%% file: main.tex
\documentclass[headings=standardclasses]{scrartcl}

\usepackage{geometry}
\usepackage[utf8]{inputenc}
\usepackage[T1]{fontenc}
\usepackage[onehalfspacing]{setspace}
\usepackage[charter]{mathdesign}
\usepackage{microtype}
\usepackage{amsmath}
\usepackage{amsthm}
\usepackage{tabulary}
\usepackage{longtable}
\usepackage{booktabs} 
\usepackage{diagbox}
\usepackage{ragged2e}

\usepackage{siunitx}
\usepackage{bm}
\usepackage[pdftex]{graphicx}
\usepackage{rotating}
\usepackage{pdfpages}
\usepackage{pdflscape}
\usepackage{color}
\usepackage{array}
\newcolumntype{L}[1]{>{\raggedright\let\newline\\\arraybackslash\hspace{0pt}}m{#1}}
\usepackage[font=bf]{caption}
\usepackage{subcaption}
\usepackage{pgfplots}
\usepackage{pgfplotstable}
\usepackage{float}
\usepackage[title]{appendix}
\pgfplotsset{compat=1.9}
\usepackage{url}
\usepackage[section]{placeins}
\usepackage{natbib}
\usepackage{hyperref}
\hypersetup{
	colorlinks   = true,    
	urlcolor     = blue,    
	linkcolor    = blue,    
	citecolor    = blue      
}
\usepackage{algorithm}
\usepackage{algpseudocode}

\makeatletter
\renewcommand*\env@matrix[1][c]{\hskip -\arraycolsep
  \let\@ifnextchar\new@ifnextchar
  \array{*\c@MaxMatrixCols #1}}
\makeatother

\makeatletter
\newcommand\notsotiny{\@setfontsize\notsotiny\@vipt\@viipt}
\makeatother

\usepackage{xcolor}
\usepackage{soul}

\usepackage{verbatim}

\begin{document}
\thispagestyle{empty}
\begin{spacing}{1.2}
\begin{center}
\huge \textbf{A Data Fusion Approach for Ride-sourcing Demand Estimation: A Discrete Choice Model 
with Sampling and Endogeneity Corrections} \\
\vspace{\baselineskip}
\end{center}
\begin{flushleft}
\normalsize
14 October 2022 \\
\vspace{\baselineskip}
\textsc{Rico Krueger} (corresponding author) \\
Department of Technology, Management and Economics \\
Technical University of Denmark (DTU), Denmark \\
rickr@dtu.dk \\
\vspace{\baselineskip}
\textsc{Michel Bierlaire} \\
Transport and Mobility Laboratory \\
Ecole Polytechnique F\'{e}d\'{e}rale de Lausanne, Switzerland \\
michel.bierlaire@epfl.ch \\
\vspace{\baselineskip}
\textsc{Prateek Bansal} \\
Department of Civil and Environmental Engineering\\
National University of Singapore, Singapore \\
prateekb@nus.edu.sg  \\
\end{flushleft}
\end{spacing}

\newpage
\thispagestyle{empty}
\section*{Abstract}

Ride-sourcing services offered by companies like Uber and Didi have grown rapidly in the last decade. Understanding the demand for these services is essential for planning and managing modern transportation systems. Existing studies develop statistical models for ride-sourcing demand estimation at an aggregate level due to limited data availability. These models lack foundations in microeconomic theory, ignore competition of ride-sourcing with other travel modes, and cannot be seamlessly integrated into existing individual-level (disaggregate) activity-based models to evaluate system-level impacts of ride-sourcing services. In this paper, we present and apply an approach for estimating ride-sourcing demand at a disaggregate level using discrete choice models and multiple data sources. We first construct a sample of trip-based mode choices in Chicago, USA by enriching household travel survey with publicly available ride-sourcing and taxi trip records. We then formulate a multivariate extreme value-based discrete choice with sampling and endogeneity corrections to account for the construction of the estimation sample from multiple data sources and endogeneity biases arising from supply-side constraints and surge pricing mechanisms in ride-sourcing systems. Our analysis of the constructed dataset reveals insights into the influence of various socio-economic, land use and built environment features on ride-sourcing demand. We also derive elasticities of ride-sourcing demand relative to travel cost and time. Finally, we illustrate how the developed model can be employed to quantify the welfare implications of ride-sourcing policies and regulations such as terminating certain types of services and introducing ride-sourcing taxes. \\
\\
\emph{Keywords:} ride-sourcing, big data, mode choice, endogeneity, travel demand.

\newpage

\section{Introduction} \label{sec:intro}

Ride-sourcing services like Uber, Lyft, Didi and Grab have expanded rapidly in the last decade and have attracted considerable ridership in many metropolitan areas worldwide \citep[see][]{goletz2021ride}. Ride-sourcing is a disruptive transport mode with positive (provision of convenient, affordable on-demand transportation options) and negative (congestion, pollution, increased vehicle kilometres travelled, possible cannibalisation of public transport demand) impacts on transport systems \citep[see][]{tirachini2020ride}. To realise their advantages and inhibit their disadvantages, ride-sourcing services need to be planned, regulated and managed \citep{goletz2021ride, tirachini2020ride}. To that end, a rigorous understanding of ride-sourcing demand is essential. Specifically, it is crucial to i) explain the characteristics of ride-sourcing demand, ii) analyse the interaction of ride-sourcing with other transport modes and iii) quantify the welfare implications of introducing ride-sourcing services or amending operational policies. 
To provide actionable, evidence-based decision support, ride-sourcing demand analysis calls for i) powerful methods to leverage datasets with varying disaggregation and resolution and ii) comprehensive datasets with user-level preferences and vehicle-level operations at an urban scale, yet both are currently found wanting. 

In terms of methods, ride-sourcing demand analysis is currently dominated by approaches without adequate foundations in microeconomic theory. Aggregate models such as regression models for count and continuous data \citep[e.g.][]{ghaffar2020modeling, marquet2020spatial} estimate a statistical relationship between realised aggregate demand and aggregate explanatory variables. Ordered outcome models for explaining ride-sourcing use at the individual level \citep[e.g.][]{alemi2019drives, von2021exploring} infer structural relationships between demand and individual-specific attributes. These methods disregard that ride-sourcing demand arises at the disaggregate level in the form of a mode choice involving trade-offs between various alternative-specific attributes (e.g. travel cost, time, reliability and safety).

Most applications of these statistical methods are driven by limited data availability. Commonly considered data sources exhibit significant weaknesses when they are analysed in isolation. Household travel surveys have a broad geographical coverage. However, they typically only include a small number of ride-sourcing trips, which precludes a thorough analysis of ride-sourcing demand. In principle, discrete choice experiments (DCEs) allow for a detailed analysis of ride-sourcing demand. However, data collected via DCEs may exhibit hypothetical biases. Also, DCEs are typically not repeated over time due to financial and logistical constraints. Recently, ride-sourcing trip records have been published under data sharing agreements between ride-sourcing companies and city authorities \citep[e.g.][]{ghaffar2020modeling}. These trip records have a broad spatiotemporal coverage. However, in isolation, they cannot be used for the disaggregate analysis of ride-sourcing demand since they do not contain any information about the demand for other modes.

This research aims at improving ride-sourcing demand analysis. To that end, we present and apply an approach for estimating ride-sourcing demand using discrete choice models (DCMs) by fusing multiple data sources. DCMs are well suited for analysing ride-sourcing demand due to their solid foundations in microeconomic theory. DCMs estimate structural relationships between observed travel choices and various alternative- and individual-specific attributes. Due to their structural nature, DCMs produce stable and transferable predictions, which in turn makes DCMs suitable for analysing counterfactual pricing and service configuration scenarios.

We first construct an estimation sample of trip-based mode choices in Chicago by enriching household travel survey data with publicly available ride-sourcing and taxi trip records. By fusing the two data sources, we address the problems of i) having too few ride-sourcing trip records in household travel survey data and ii) having no information about other modes in ride-sourcing trip records. However, constructing an estimation sample from two revealed preference data sources creates two challenges in developing a DCM. First, a sampling correction is needed to account for the enrichment of the household travel survey data with ride-sourcing and taxi trip records. Second, the constructed mode choice dataset likely exhibits endogeneity biases, as the demand for the ride-sourcing options and their prices is simultaneously determined by supply-side constraints and surge pricing mechanisms \citep[see e.g.][]{castillo2017surge}. We address both challenges by formulating a multivariate extreme value (MEV)-based DCM with sampling and endogeneity corrections. To correct for sampling biases, we adopt a conditional maximum likelihood estimator \citep[see][]{bierlaire2020sampling} due to its efficiency properties; and to correct for endogeneity biases, we adopt the control function approach \citep{petrin2010control} due to its simplicity. Ultimately, we apply the model to the constructed mode choice dataset to analyse the demand for ride-sourcing services in Chicago. The parameter estimates of the DCM are translated into the elasticity of ridesourcing demand relative to alternative-specific attributes like price and travel time. 
We also illustrate how the estimated model can be employed to quantify the welfare implications of ride-sourcing policies and regulations such as eliminating certain types of services and introducing taxes. 

We organise the remainder of this paper as follows:
In Section~\ref{sec:lit}, we review related literature. 
In Section~\ref{sec:data}, we describe the construction of the estimation sample for the empirical application. 
In Section~\ref{sec:model}, we present the general formulation of the econometric model.
In Section~\ref{sec:specification}, we explain the model specification considered in the empirical application. 
In Section~\ref{sec:results}, we discuss the results of the empirical application, and in Section~\ref{sec:welfare}, we present the welfare analysis.
Finally, we conclude in Section~\ref{sec:conclusion}.

\section{Related literature} \label{sec:lit}

The literature on ride-sourcing demand analysis evolves rapidly. In Table~\ref{tab:litreview} in Appendix~\ref{app:lit_review}, we present an overview of recently published ride-sourcing demand analysis studies. For reviews of earlier studies, the reader is directed to \citet{tirachini2020ride} and \citet{wang2019ridesourcing}. The studies enumerated in Table~\ref{tab:litreview} can be subsumed under five topics: 
\begin{enumerate}
\item travel modes (mostly public transit) that are complemented or substituted by ride-sourcing services; impact of emerging on-demand mobility services on vehicle ownership;
\item association of built-environment, socio-demographics, weather and land use characteristics with ride-sourcing demand at a spatial level (e.g. census tract and census block groups);
\item association of attitudes, socio-demographic and economic characteristics with the usage of and preferences for ride-sourcing services;
\item determinants of preferences for the use of pooled ride-sourcing services;
\item impact of mode-specific attributes (e.g. travel time and wait time) on the demand for these services in the multi-modal transport system.
\end{enumerate}

The studies use mainly three types of data (see column ``data type'' in Table \ref{tab:litreview}). First, trip records from ride-sourcing companies are merged with supplementary data about land use, weather and census tract attributes. These studies focus predominantly on the first two of the five topics enumerated above; only a few studies focus on the fourth topic. Second, household travel surveys with information about individuals' travel patterns, socio-demographic characteristics and attitudes are considered for exploring the third topic; a handful of studies also focus on topics one and four. Third, DCEs are employed to investigate topics three to five. 

In terms of methods, most studies that consider the first data type aggregate trips across space and time and then rely on geographically weighted and spatially lagged count or continuous data models with autoregressive structure or panel effects. A few studies use off-the-shelf machine learning algorithms such as random forest or gradient boosting decision trees. Studies considering household travel survey data use multinomial and ordered logit models. Several studies also develop joint models of continuous, count and ordered variables (i.e.\ generalised heterogeneous data models). Structural equation models are also used to analyse the relationship between ride-sourcing demand and latent attitudes. Finally, studies collecting data through DCEs naturally employ DCMs such as nested, latent class, error component or mixed logit models.

Only a few studies use DCMs and revealed preference data to analyse ride-sourcing demand. \citet{habib2019mode} considers revealed preference data from a household travel survey to investigate ride-sourcing demand using a semi-compensatory choice model with probabilistic choice set formation. The study finds that ride-sourcing demand mostly complements the demand for driving and transit and substitutes taxi demand. Furthermore, the probability of considering ride-sourcing varies by age, whereby young people are more likely to consider ride-sourcing, and older people are more likely to consider taxis. 
\citet{lam2021geography} considered revealed preference data constructed from ride-sourcing trip records, field data and API queries to analyse ride-sourcing demand in New York City. The authors employ an aggregate logit model for market share data. The model includes endogeneity corrections for price and wait times. The study finds that the distribution of ride-sourcing benefits varies substantially across space, with low accessibility areas experiencing comparatively higher benefits.

In summary, household travel surveys and trip records have been used in isolation. 
Both data sources exhibit significant weaknesses when used in isolation: Household travel surveys contain insufficient information about ride-sourcing demand; trip records cannot be used for disaggregate demand analysis, as they do not include information about individual-level preferences for other travel modes.
This current study contributes to the literature with a DCM for disaggregate demand analysis of ride-sourcing services by fusing both datasets and addressing potential endogeneity issues. This data fusion framework leverages the richness of both data types while addressing the shortcomings of analysing them in isolation. We also control for demographics, transit accessibility, parking cost, land use characteristics, pedestrian friendliness, and weather conditions in the DCM, which may affect demand for ridesourcing services. The developed model can be used as a direct input to activity-based travel demand forecasting models to quantify the short- and long-term impact of policies and regulations related to ridesourcing services on the multi-modal transport system.   

Finally, our study is also related to the literature on endogeneity and discrete choice analysis in various other applications, including but not limited to 
consumer choice \citep{petrin2010control}, 
residential location choice \citep{guevara2012change}, 
airline itinerary choice \citep{lurkin2017accounting}
and parking choice \citep{gopalakrishnan2020combining}.

\section{Data} \label{sec:data}

\begin{figure}[H]
\centering
\includegraphics[width = 0.6 \textwidth]{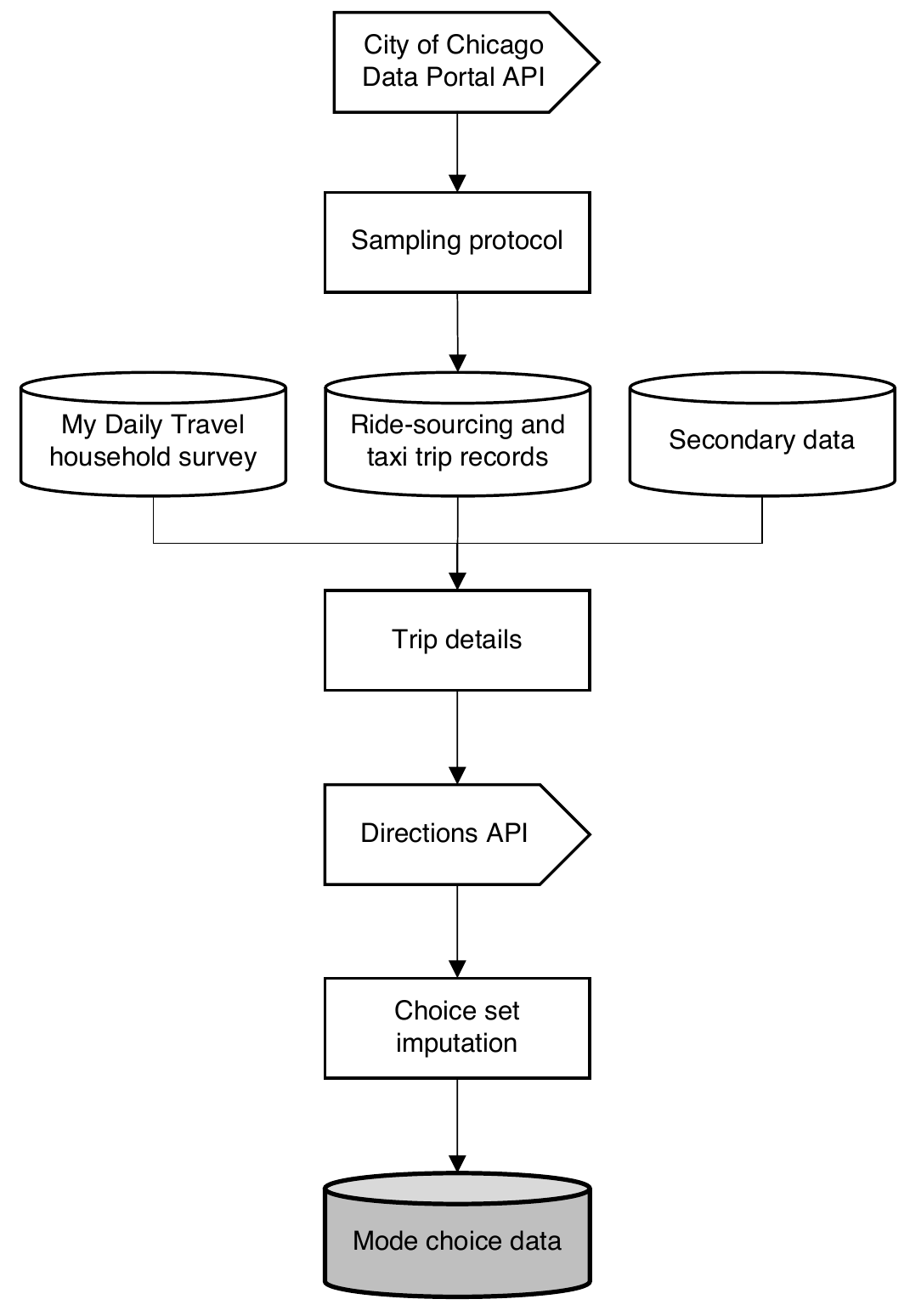}
\caption{Dataset construction} \label{fig:data_construction}
\end{figure}

We construct an estimation sample of trip-based mode choices in Chicago from November 2018 to May 2019, following the process visualised in Figure~\ref{fig:data_construction}. 
In what follows, we describe the construction of the estimation sample in detail.

\subsection{Primary data sources}

Trip records for the construction of the estimation sample are gathered from two primary sources, namely i) ride-sourcing and taxi trip records provided on the City of Chicago Data Portal\footnote{\url{https://data.cityofchicago.org}} and ii) the My Daily Travel household survey. 

The City of Chicago Data Portal provides access to records of all ride-sourcing and taxi trips that transportation networking providers and taxi companies have reported to the City of Chicago for regulatory purposes since November 2018 and January 2013, respectively. 
The attributes of the trips are temporally and spatially aggregated to prevent a re-identification of individual trips. 
Each trip record includes information about the trip start and end times rounded to the nearest 15 minutes, the pick-up and drop-off community areas as well as the fare amount. The pick-up and drop-off census tracts are also available if at least three trips started or ended in the relevant census tract in the relevant 15-minute period. 
For ride-sourcing trips, it is also known if a pooled trip was requested. Thus, solo and pooled ride-sourcing trips can be distinguished. 

The My Daily Travel household survey is a large-scale household travel survey that was conducted by the Chicago Metropolitan Agency for Planning (CMAP) between August 2018 and May 2019. 
The survey collected information about the daily travel behaviour of a representative sample of more than 12,000 households in North-eastern Illinois. 
More information about the survey is available in \citet{westat2020my}. 
The collected data include trip records with information about the chosen transport mode (car, transit, bike, walking, taxi, solo-ride-sourcing or pooled ride-sourcing), trip start and end times as well as origin and destination census tracts. 

For the construction of the estimation sample, we exploit the temporal and spatial overlap of the ride-sourcing and taxi trip records from the City of Chicago Data Portal and the My Daily Travel household survey. 
Consequently, we limit our analysis to trip records produced between 1 November 2018 and 3 May 2019.
Since we are interested in understanding ride-sourcing use in the context of general travel demand, we only consider trips on weekdays between 5:00 and 23:00.
Furthermore, to make it possible to impute the attributes of non-chosen alternatives, we restrict our analysis to trips for which the reported start and end locations are distinct.
The start and end points of a trip are given by the centroids of the origin and destination census tracts or community areas.
We exclude trips that start or end outside of the City of Chicago. 

After applying these inclusion criteria, we are left with 12,593 trip records from the My Daily Travel household travel survey. 
For our analysis, we consider all trips from the My Daily Travel household survey that satisfy the inclusion criteria. 
18,784,655 solo ride-sourcing, 7,290,921 pooled ride-sourcing and 3,821,709 taxi trips from the City of Chicago Data Portal satisfy the inclusion criteria. 

\begin{figure}[H]
\centering
\includegraphics[width = 0.48 \textwidth]{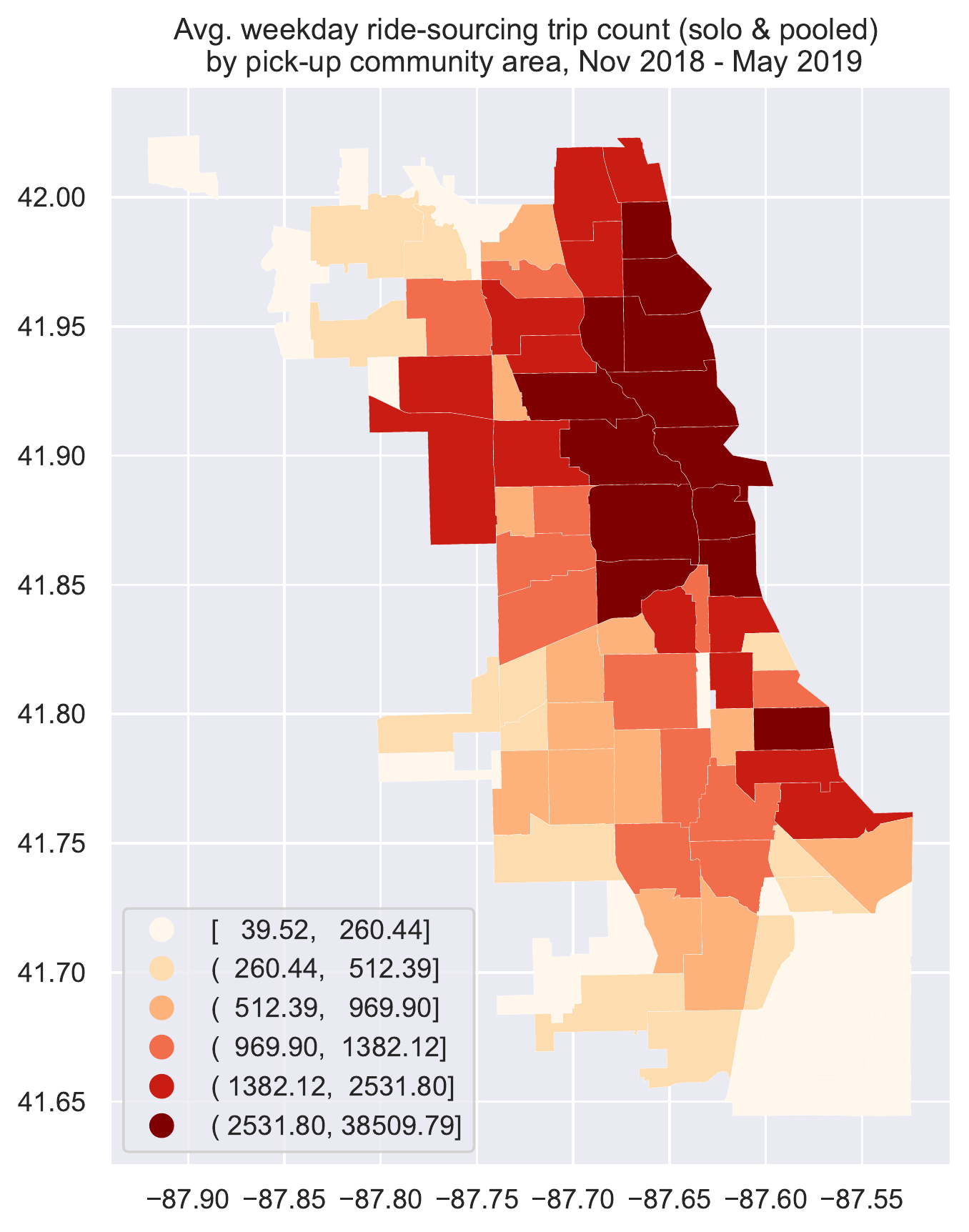}
\includegraphics[width = 0.48 \textwidth]{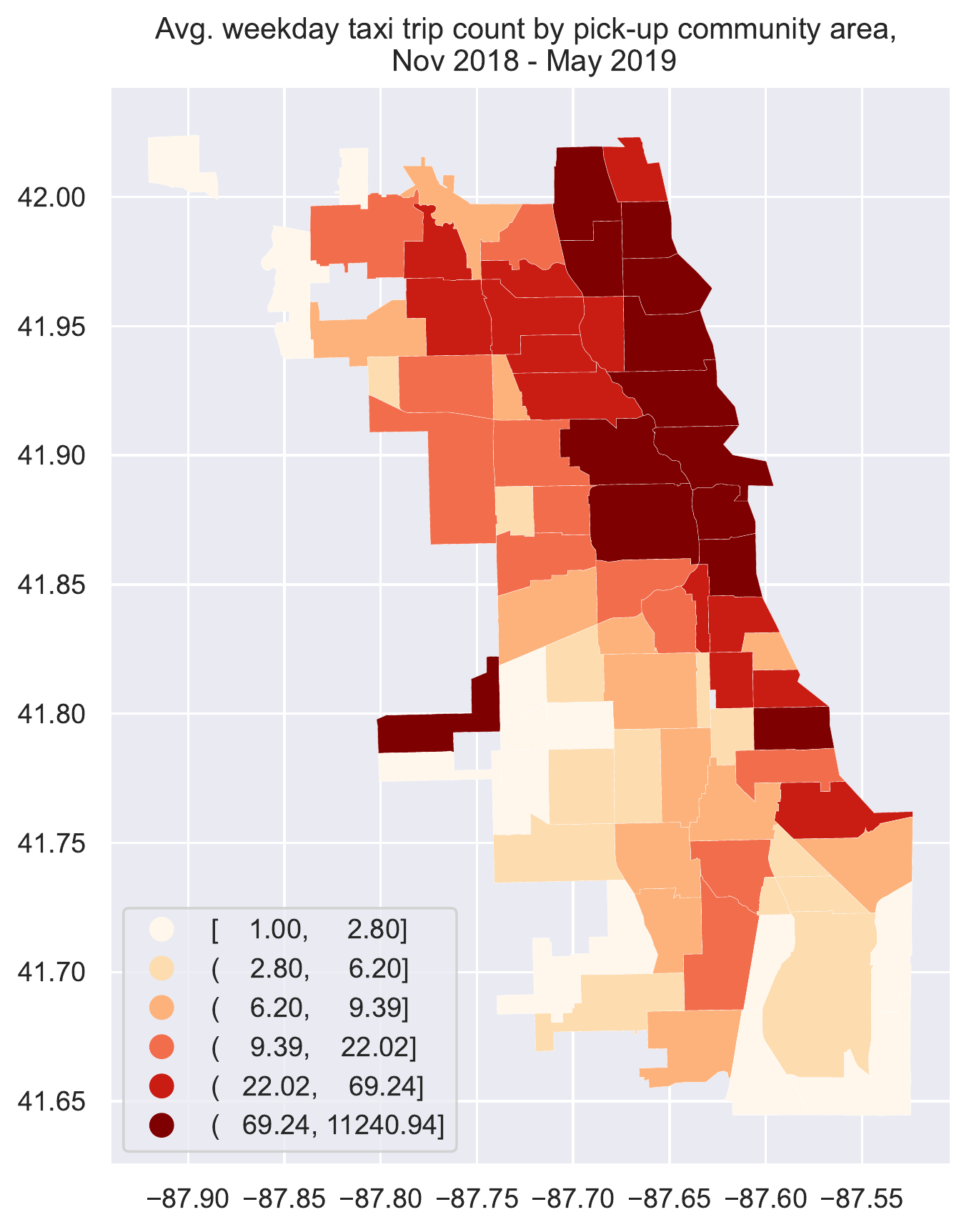}
\caption{Average weekday ride-sourcing and taxi trip counts by pick-up community area} \label{fig:avg_ride-sourcing_taxi_trip_counts}
\end{figure}

\begin{figure}[H]
\centering
\includegraphics[width = 0.48 \textwidth]{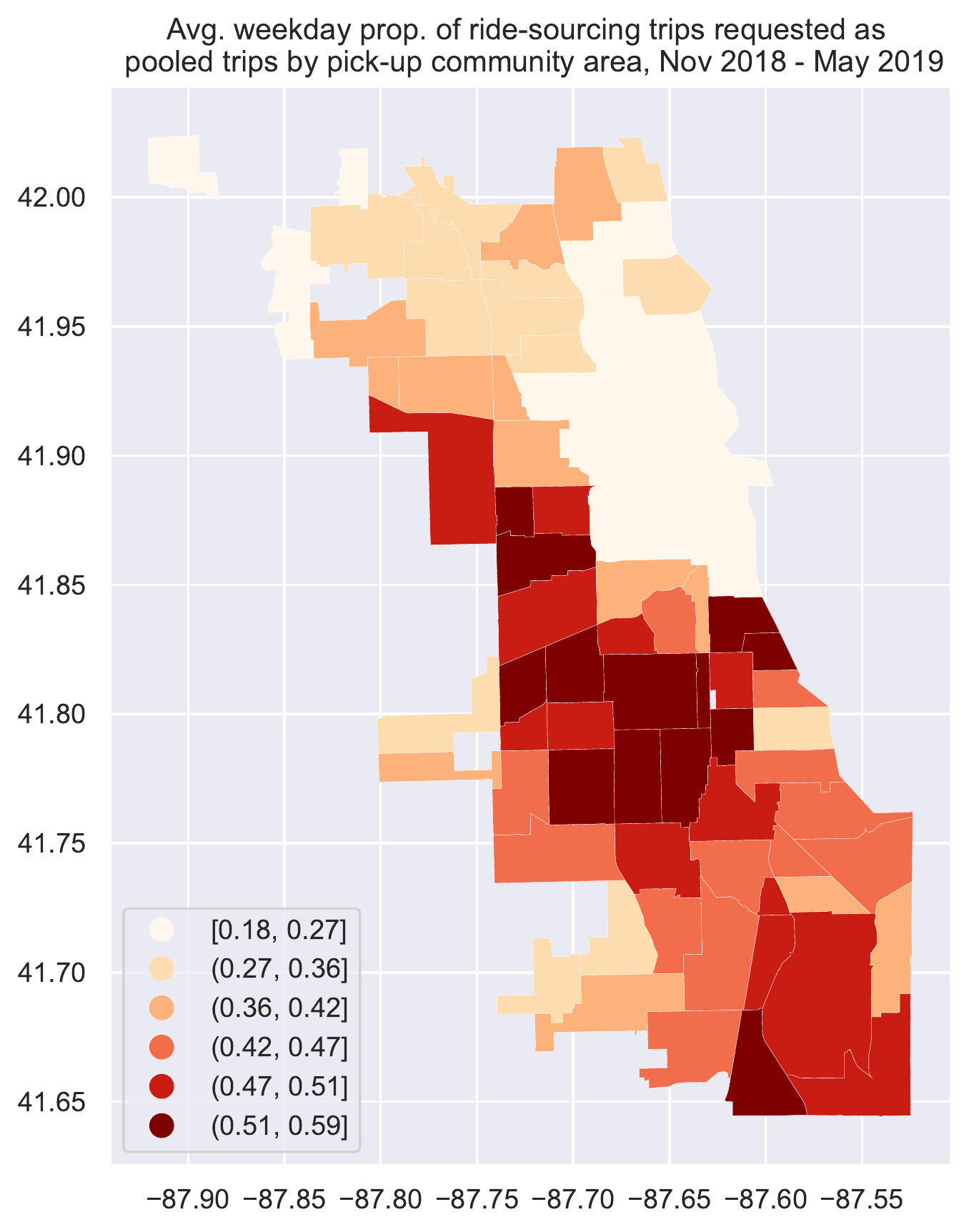}
\caption{Average proportion of ride-sourcing trips requested as pooled trips by pick-up community area} \label{fig:avg_pooled}
\end{figure}

\begin{figure}[H]
\centering
\includegraphics[width = 0.7 \textwidth]{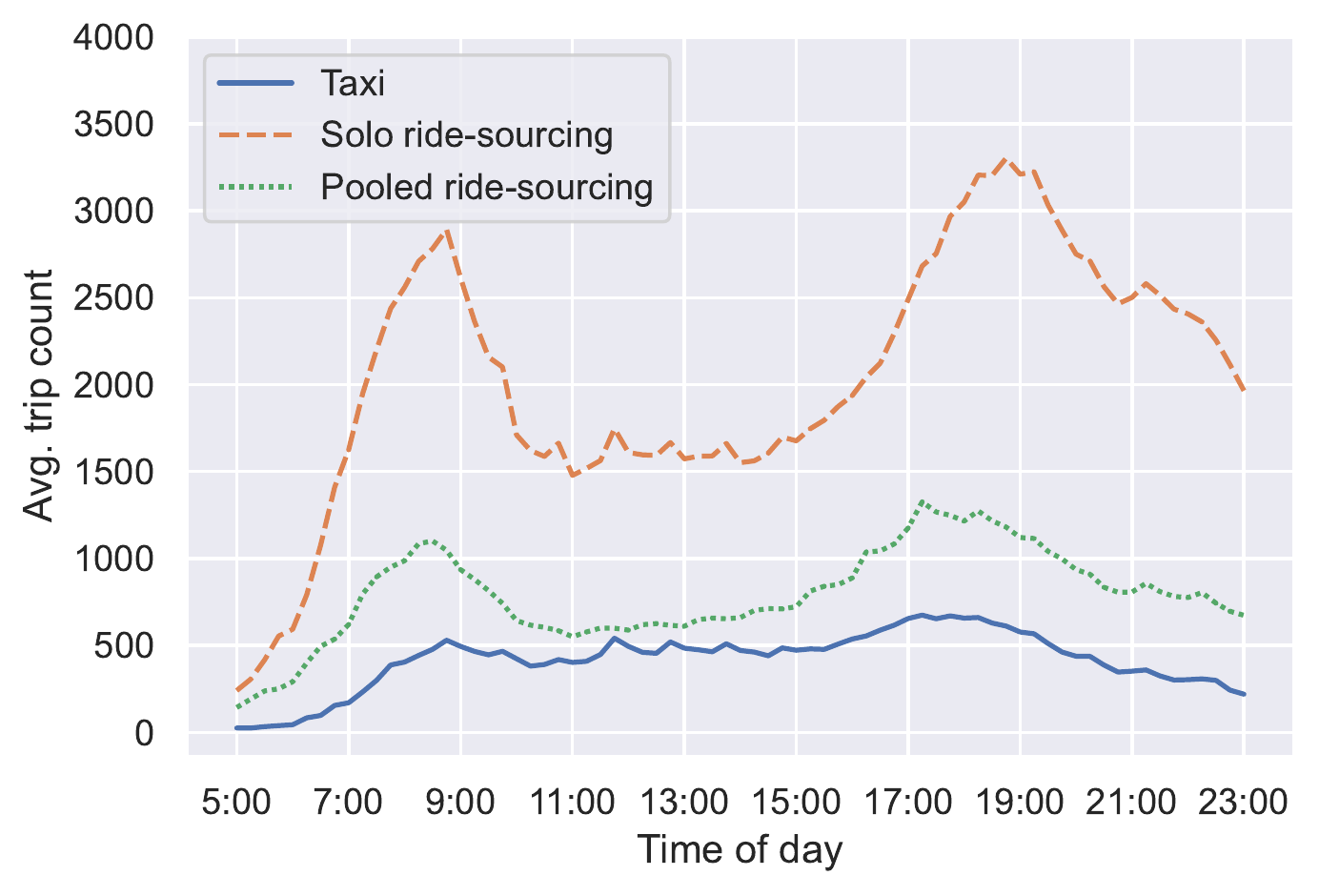}
\caption{Average ride-sourcing and taxi trip count by time of day} \label{fig:avg_trip_count_time_of_day}
\end{figure}

We briefly describe the ride-sourcing and taxi trip records that meet the inclusion criteria.  
Figure~\ref{fig:avg_ride-sourcing_taxi_trip_counts} shows the average weekday ride-sourcing and taxi trip counts by pick-up community area.
It can be seen that the demand for ride-sourcing and taxi trips is concentrated in central zones (i.e. the Northeast) of the study area.
In addition, Figure~\ref{fig:avg_pooled} shows the average proportion of ride-sourcing trips requested as pooled trips by pick-up community area.
We observe that the proportion of ride-sourcing trips requested as pooled trips is larger in the peripheral areas of the study area. 
Finally, Figure~\ref{fig:avg_trip_count_time_of_day} visualises the average ride-sourcing and taxi trip count in the whole study area by time of day. 
Whereas the demand for taxi is relatively balanced throughout the day, the demand for solo and pooled ride-sourcing trips exhibits pronounced morning and evening peaks. 

\subsection{Sampling protocol} \label{sec:sampling_protocol}

If all eligible ride-sourcing and taxi trip records were considered, the resulting estimation sample would be too large to be processed.
Therefore, we employ a choice-based sampling strategy to select ride-sourcing and taxi trips from the City of Chicago Data Portal. 
More specifically, we randomly select 20,000 records each from the sets of eligible solo ride-sourcing, pooled ride-sourcing and taxi trips.
Table~\ref{tab:shares} shows the absolute and relative frequencies of the observed mode choices in the two subsamples and the final estimation sample. 

We use the My Daily Travel household survey to estimate the population mode share. 
In line with the sample design, the population quantity of interest is the mode share for trips between distinct census tracts in Chicago on weekdays between 5:00 and 23:00. To that end, we calculate average person trip rates of the population of Northeastern Illinois using the person-specific sampling weights provided in the My Daily Travel household survey. The calculated population mode share is shown in the column ``Population---Share'' in Table~\ref{tab:shares}. 

\begin{table}[H]
\centering
\small
\input{table_shares}

\caption{Population and sample mode choice frequencies} \label{tab:shares}
\end{table}

\subsection{Secondary data sources}

After merging the two subsamples, we supplement the resulting dataset with information from secondary data sources.
 
\begin{figure}[H]
\centering
\includegraphics[width = 0.48 \textwidth]{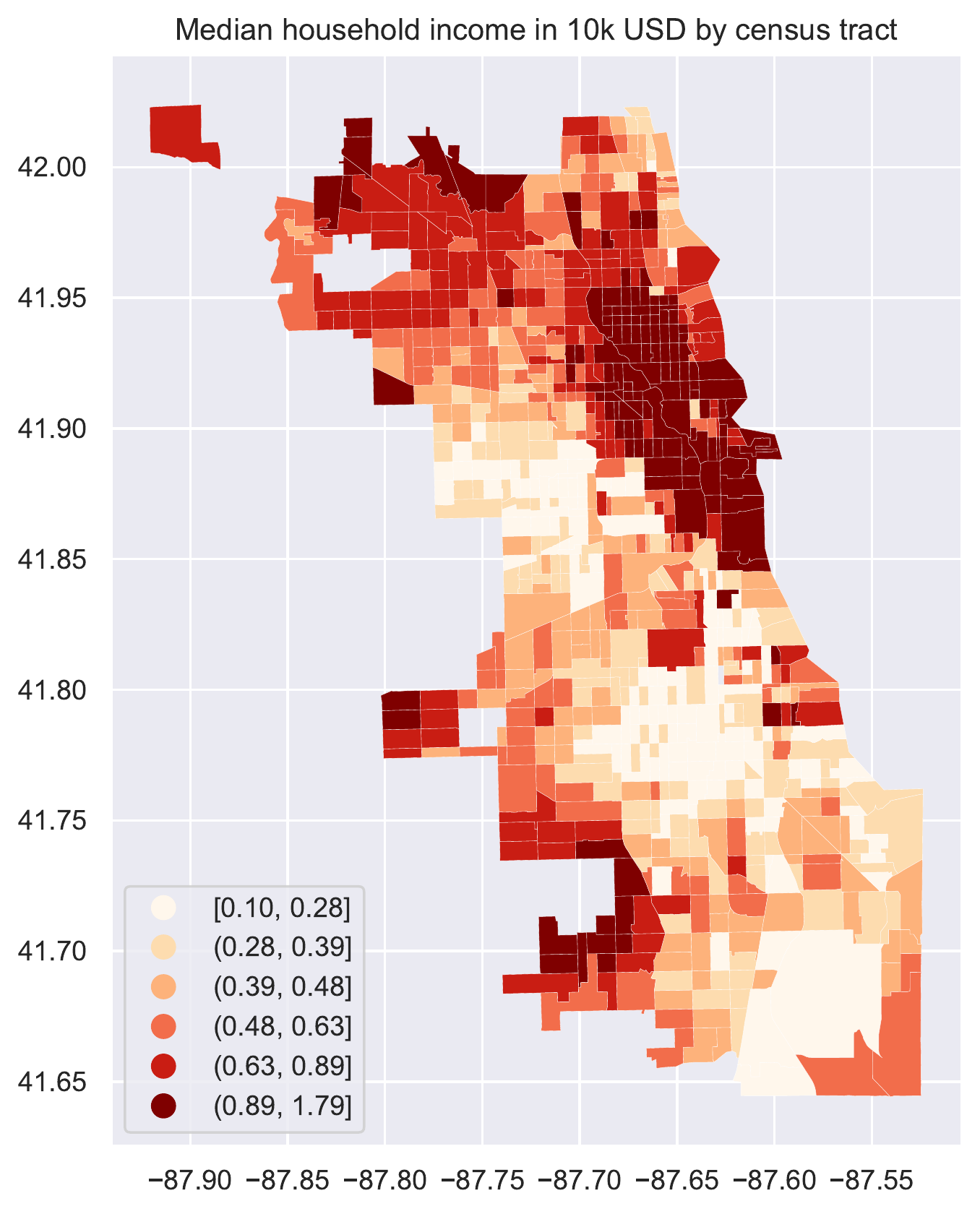}
\includegraphics[width = 0.48 \textwidth]{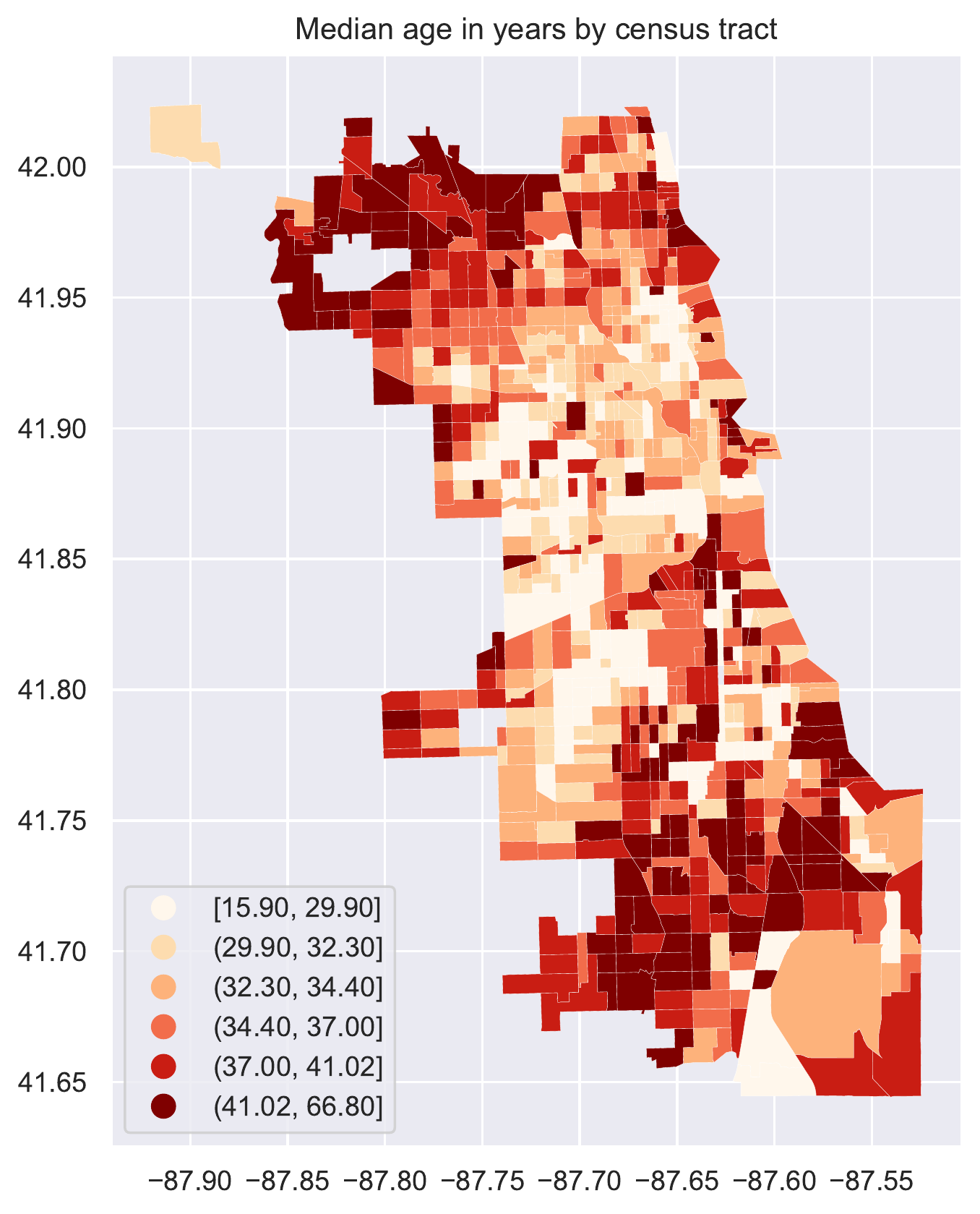}
\caption{Median household income and median age by census tract} \label{fig:income_age}
\end{figure}

The median household income and median age of each census tract are obtained from the American Community Survey \citep{uscb2021acs}. 
The spatial distributions of the two quantities are visualised in Figure~\ref{fig:income_age}.

\begin{figure}[H]
\centering
\includegraphics[width = 0.48 \textwidth]{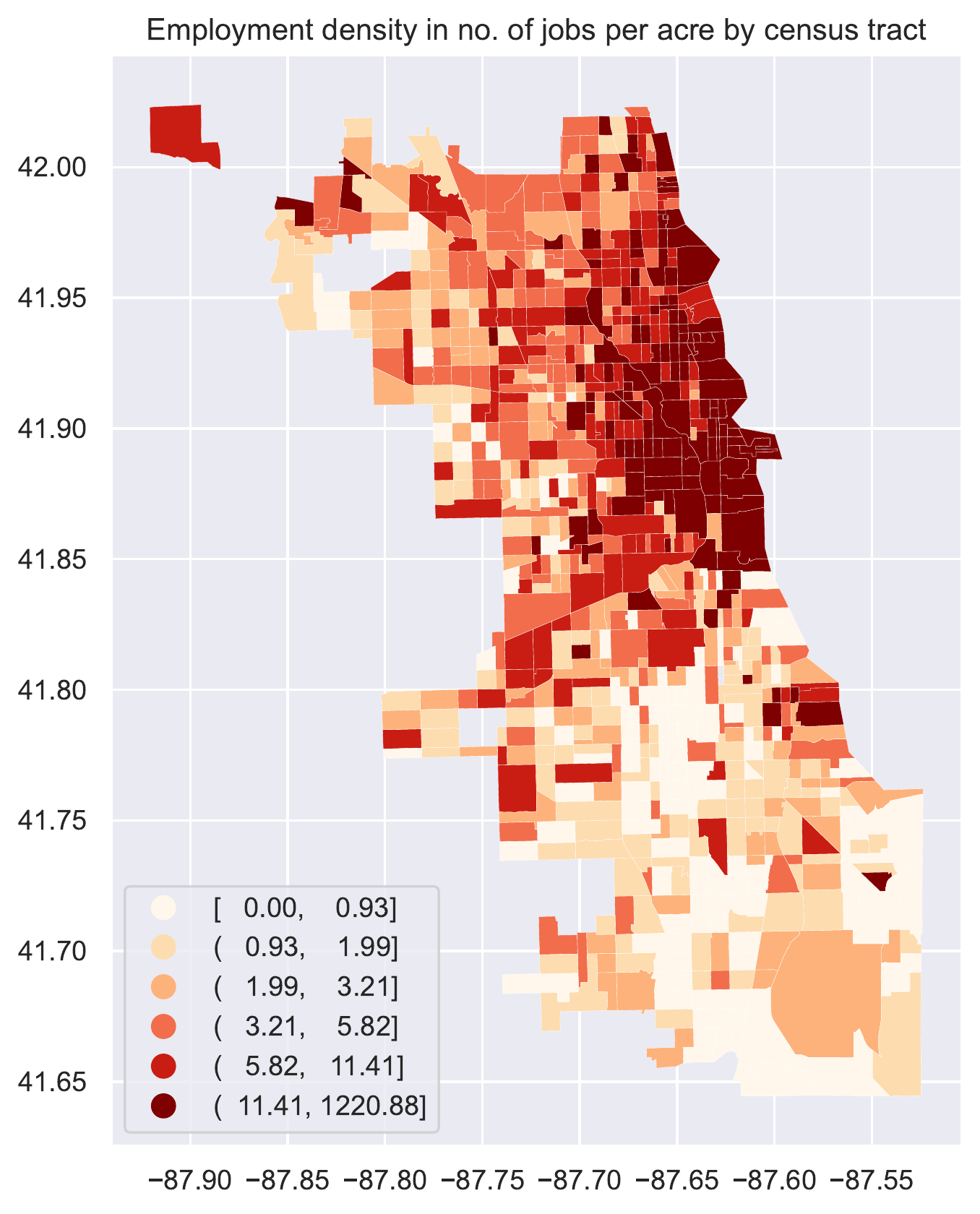}
\includegraphics[width = 0.48 \textwidth]{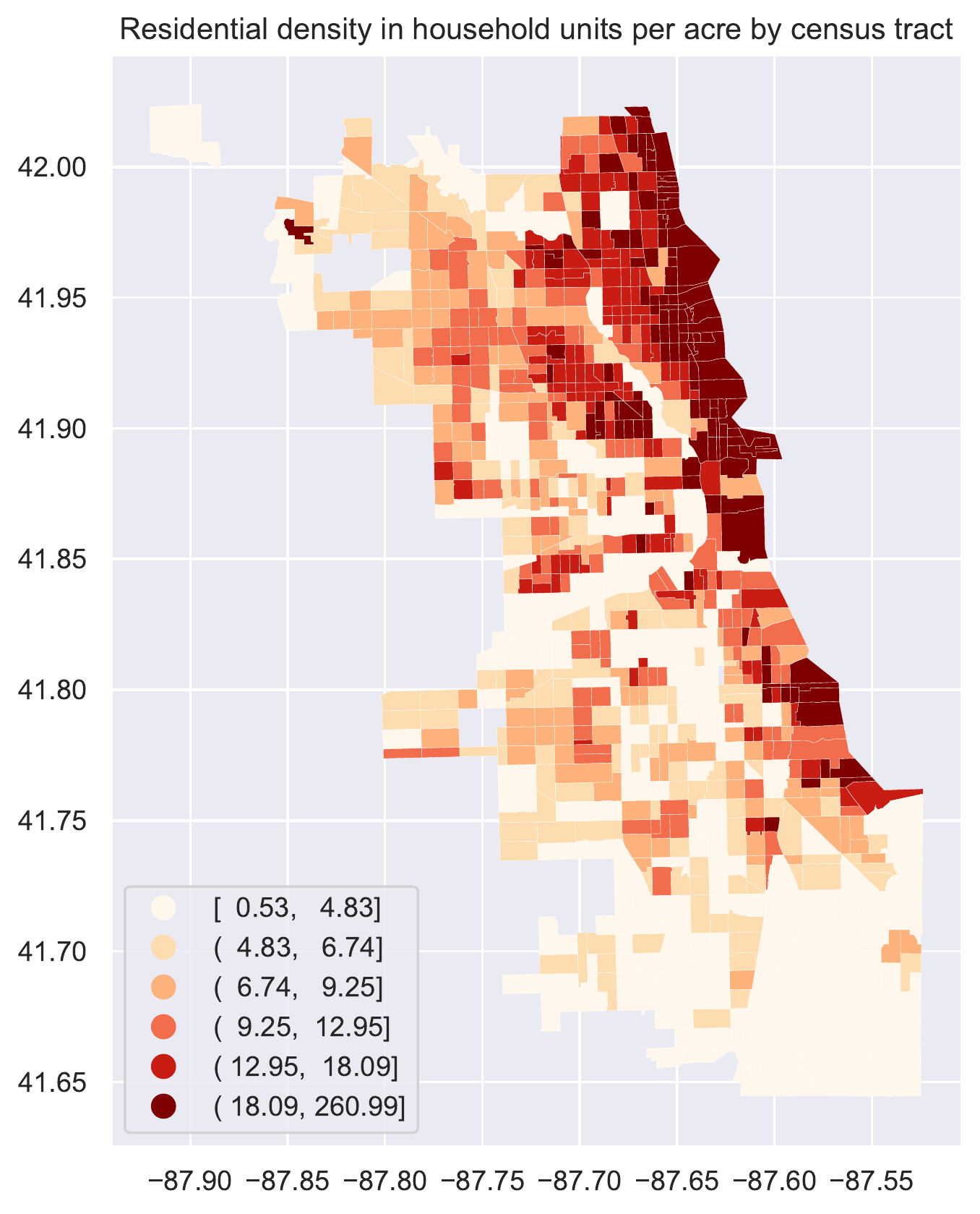}
\caption{Employment and residential densities by census tract} \label{fig:emp_res}
\end{figure}

\begin{figure}[H]
\centering
\includegraphics[width = 0.48 \textwidth]{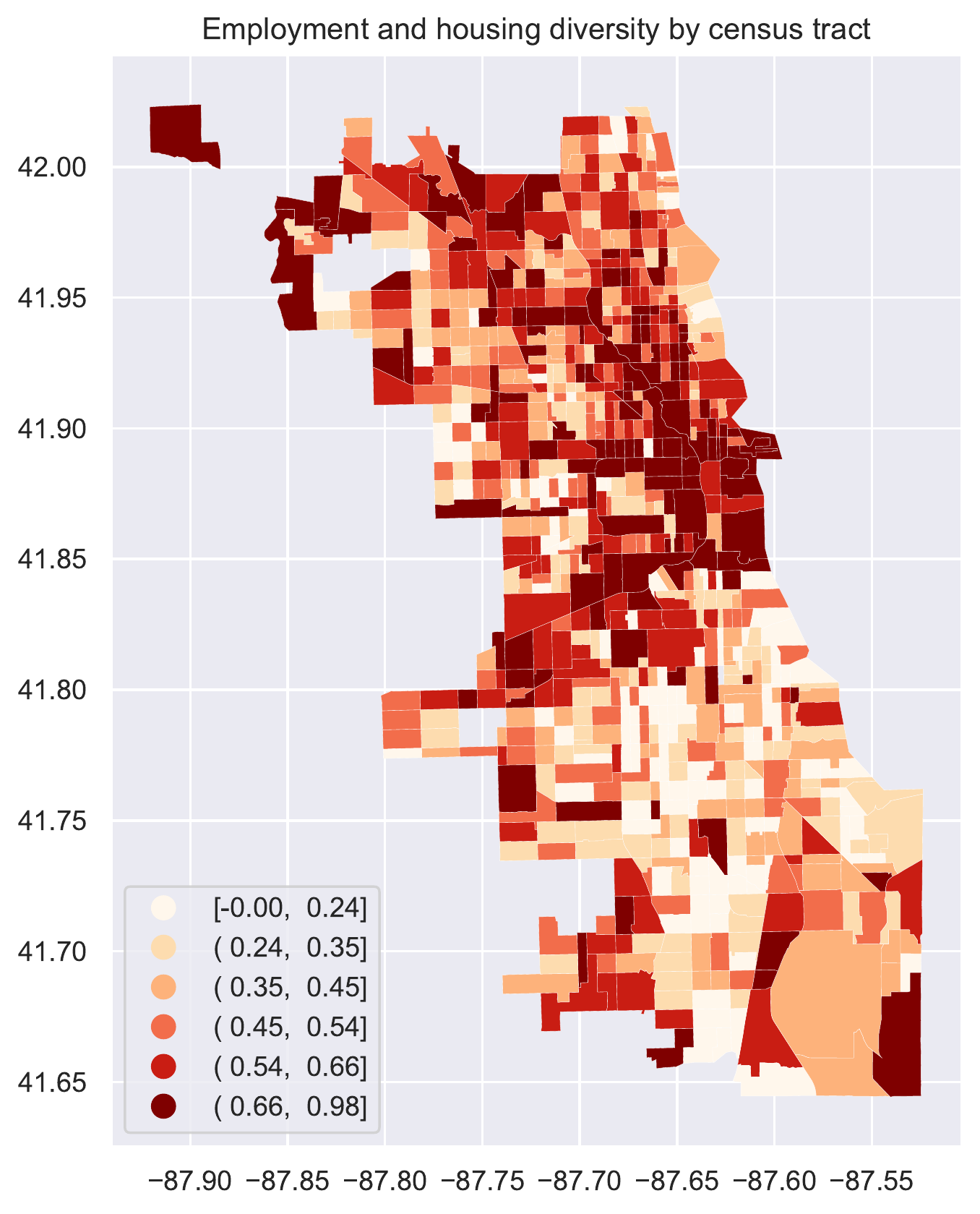}
\includegraphics[width = 0.48 \textwidth]{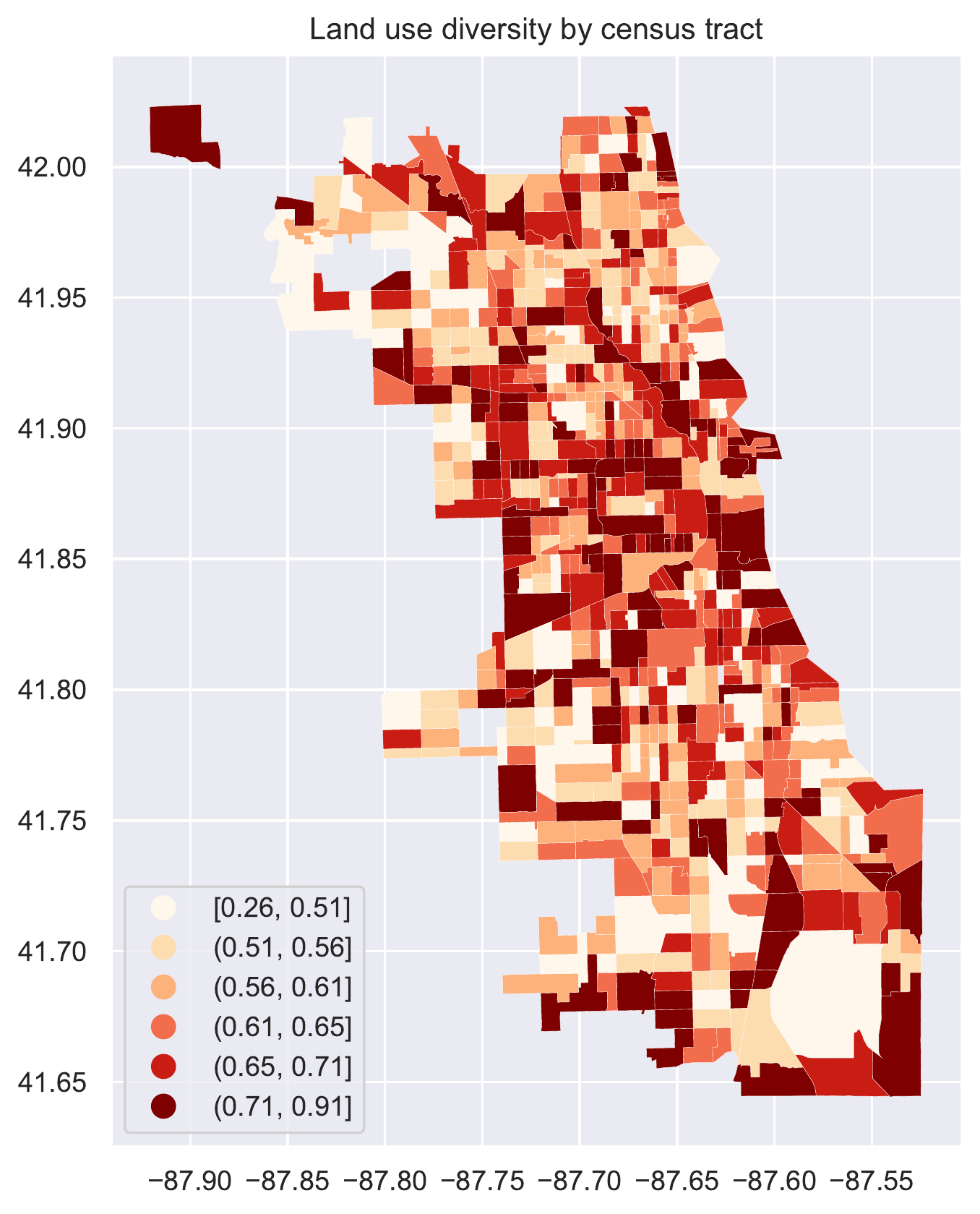}
\caption{Employment and housing diversity and land use diversity by census tract} \label{fig:land_emp_hh_div}
\end{figure}

\begin{figure}[H]
\centering
\includegraphics[width = 0.48 \textwidth]{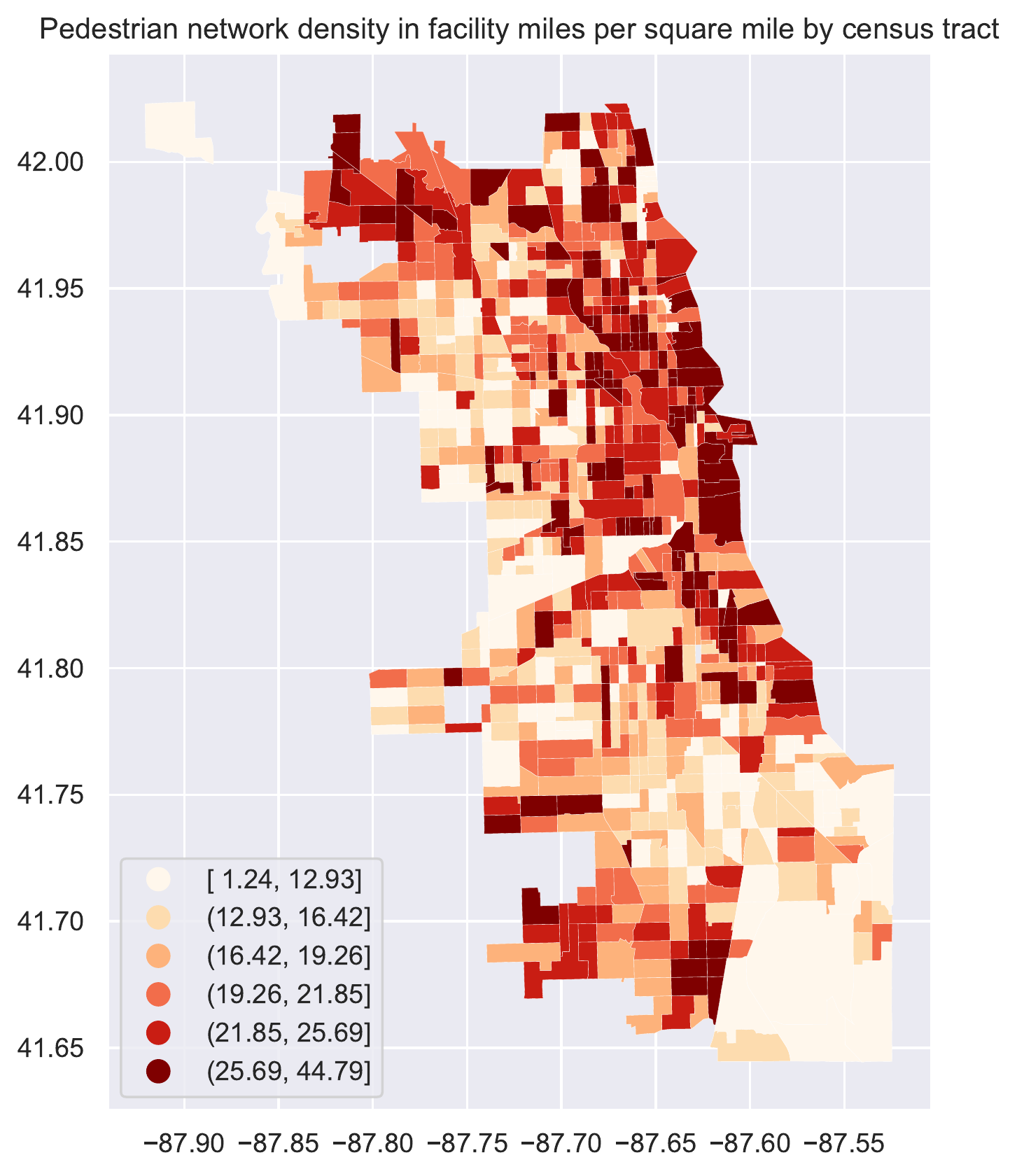}
\includegraphics[width = 0.48 \textwidth]{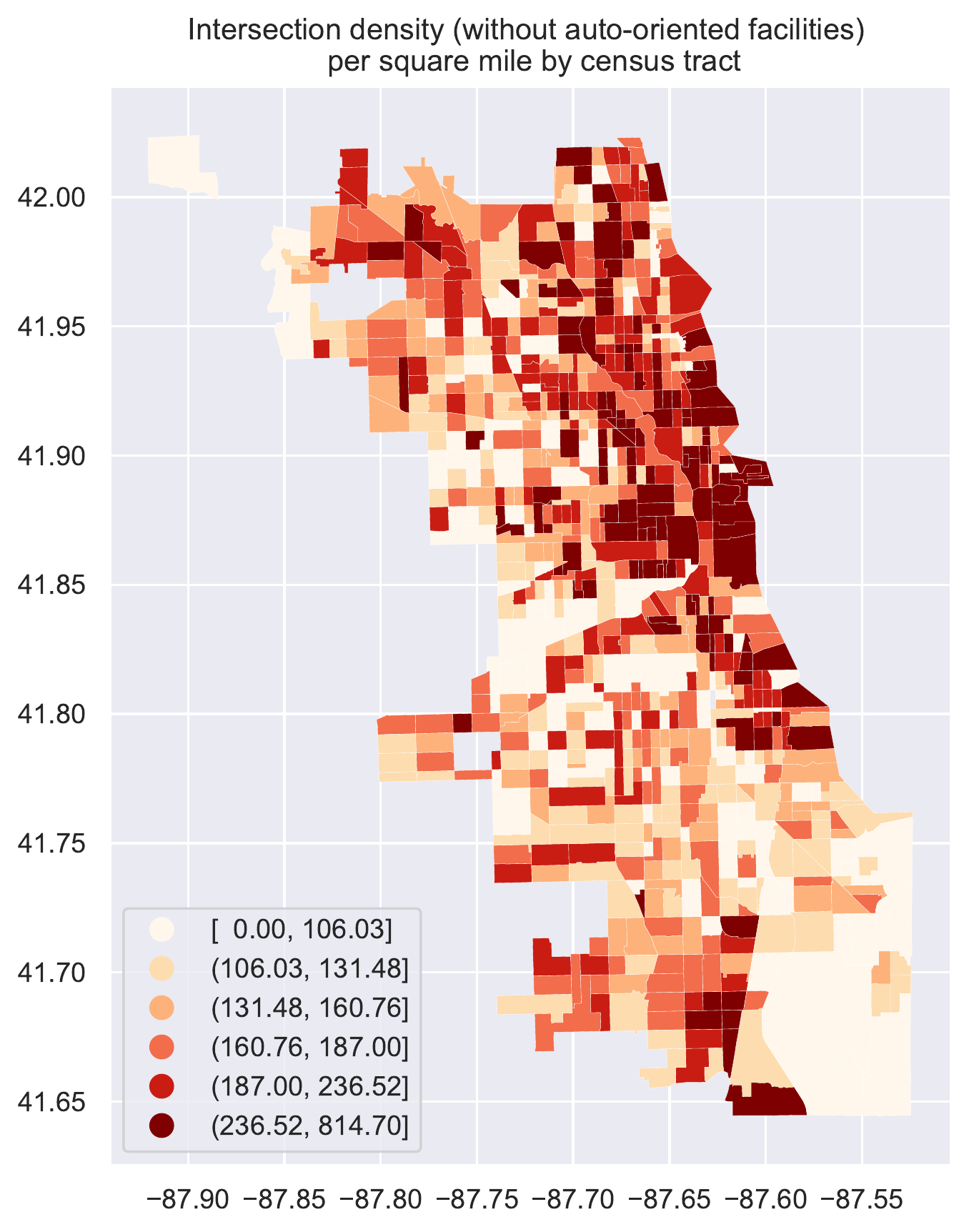}
\caption{Pedestrian network and intersection densities by census tract} \label{fig:ped_inter}
\end{figure}

\begin{figure}[H]
\centering
\includegraphics[width = 0.48 \textwidth]{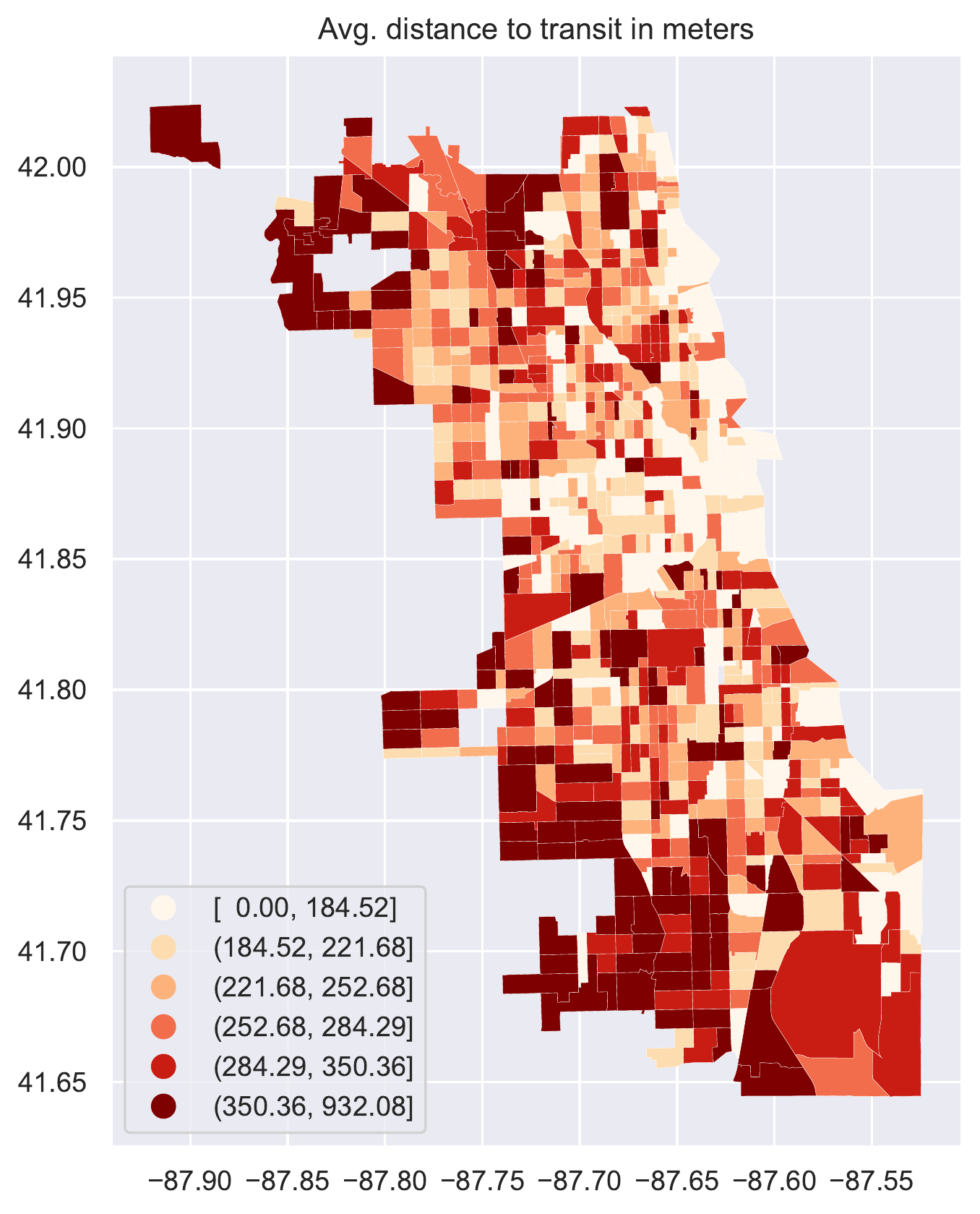}
\includegraphics[width = 0.48 \textwidth]{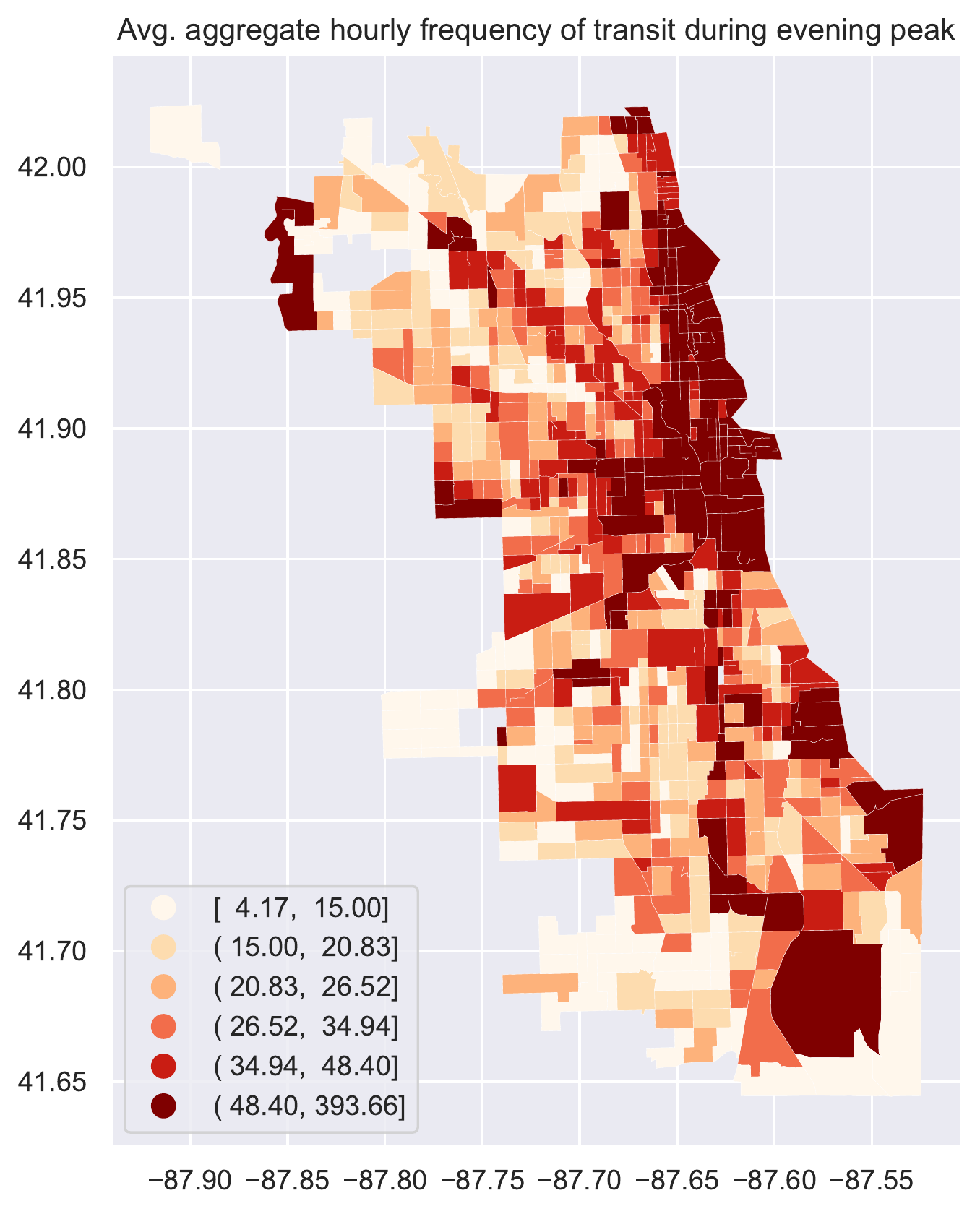}
\caption{Average proximity to transit and average transit service frequency by census tract} \label{fig:prox_freq}
\end{figure}

\begin{figure}[H]
\centering
\includegraphics[width = 0.48 \textwidth]{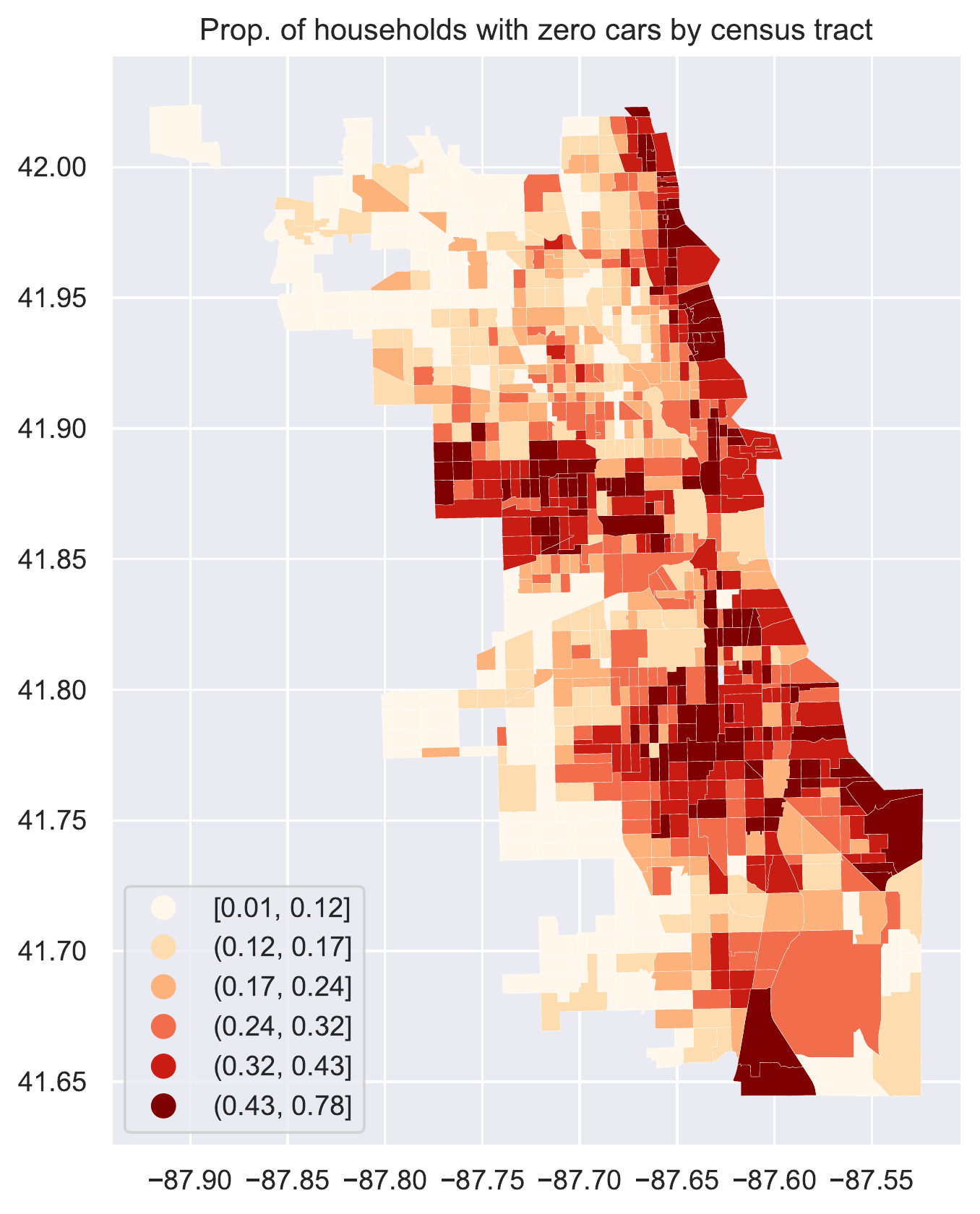}
\includegraphics[width = 0.48 \textwidth]{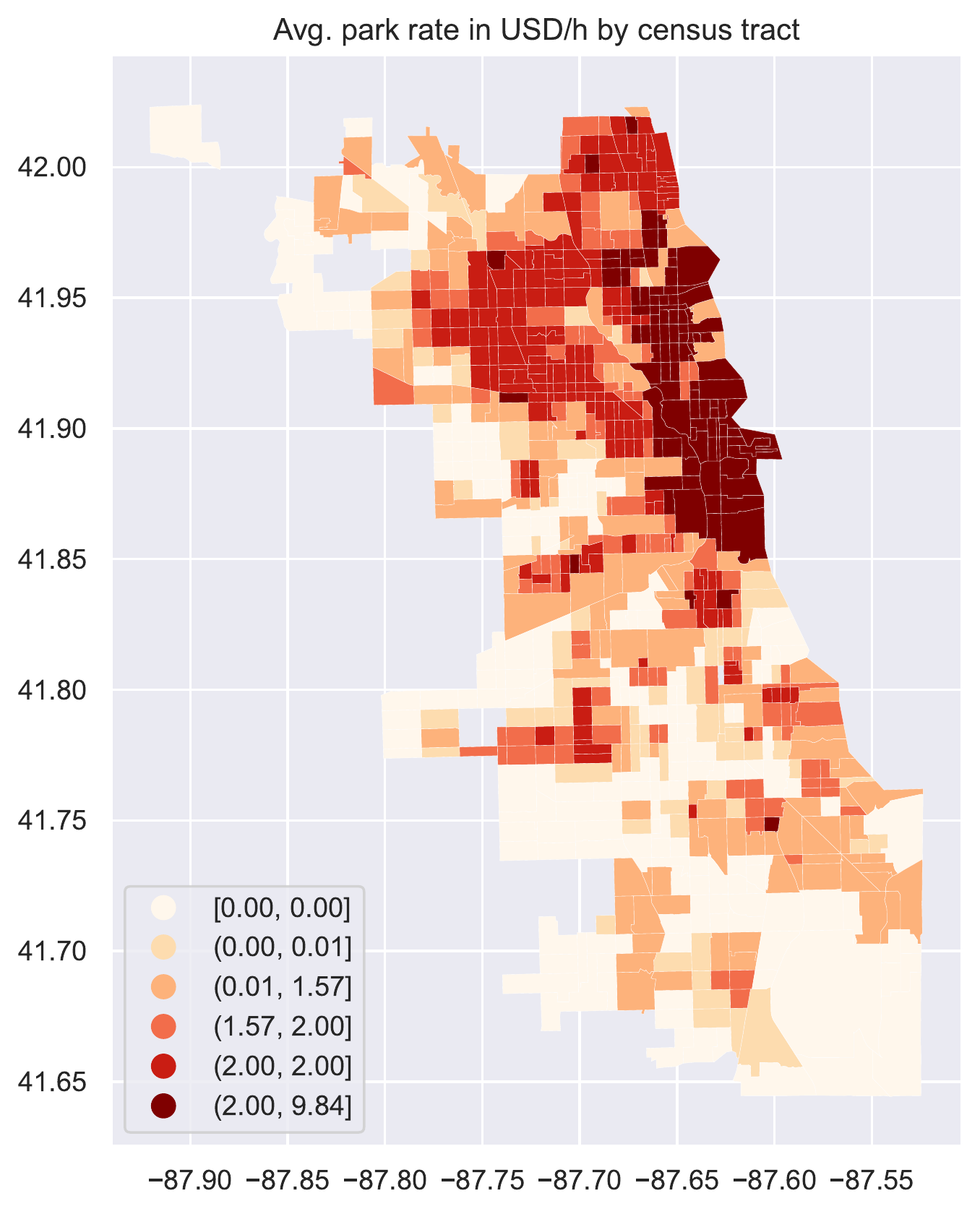}
\caption{Proportion of households with zero cars and average park rate by census tract} \label{fig:prop_zero_park}
\end{figure}

We source various census tract attributes pertaining to
employment and housing (employment density, residential density, employment and housing diversity),
pedestrian friendliness (pedestrian network density, intersection density),
transit supply (average proximity to transit, average transit service frequency)
and car ownership (proportion of households with zero cars)
from the Smart Location Database maintained by the US Environmental Protection Agency \citep{epa2021sld}. 
Employment and housing diversity is an entropy-based diversity index accounting for employment numbers in five categories (retail, office, industrial, service and entertainment) and occupied housing from the database. 
Figures~\ref{fig:income_age}--\ref{fig:prop_zero_park} show the spatial distributions of the extracted quantities.

Furthermore, information on seven land use categories (residential, commercial, institutional, industrial, transportation / communication / utilities / waste, agricultural, open space) is gathered from the 2013 CMAP Land Use Inventory \citep{cmap2015land}. 
For each census tract in the study area, we calculate an entropy-based land use diversity index of the form $D = \left ( \sum_{c \in \mathcal{C}} p_c \ln p_c \right ) \big/ \vert \mathcal{C} \vert$, where $\mathcal{C}$ denotes the set of considered land use categories, and $p_c$ is the proportion of land use classified as category $c$.
The right panel of Figure~\ref{fig:land_emp_hh_div} shows the spatial distribution of the calculated diversity index. 
 
In addition, information about park fees are sourced from the CMAP Parking Inventory \citep[see][]{ghaffar2020modeling}. The spatial distribution of the average hourly park rate is shown in the right panel of Figure~\ref{fig:prop_zero_park}. 

\begin{figure}[H]
\centering
\includegraphics[width = 0.7 \textwidth]{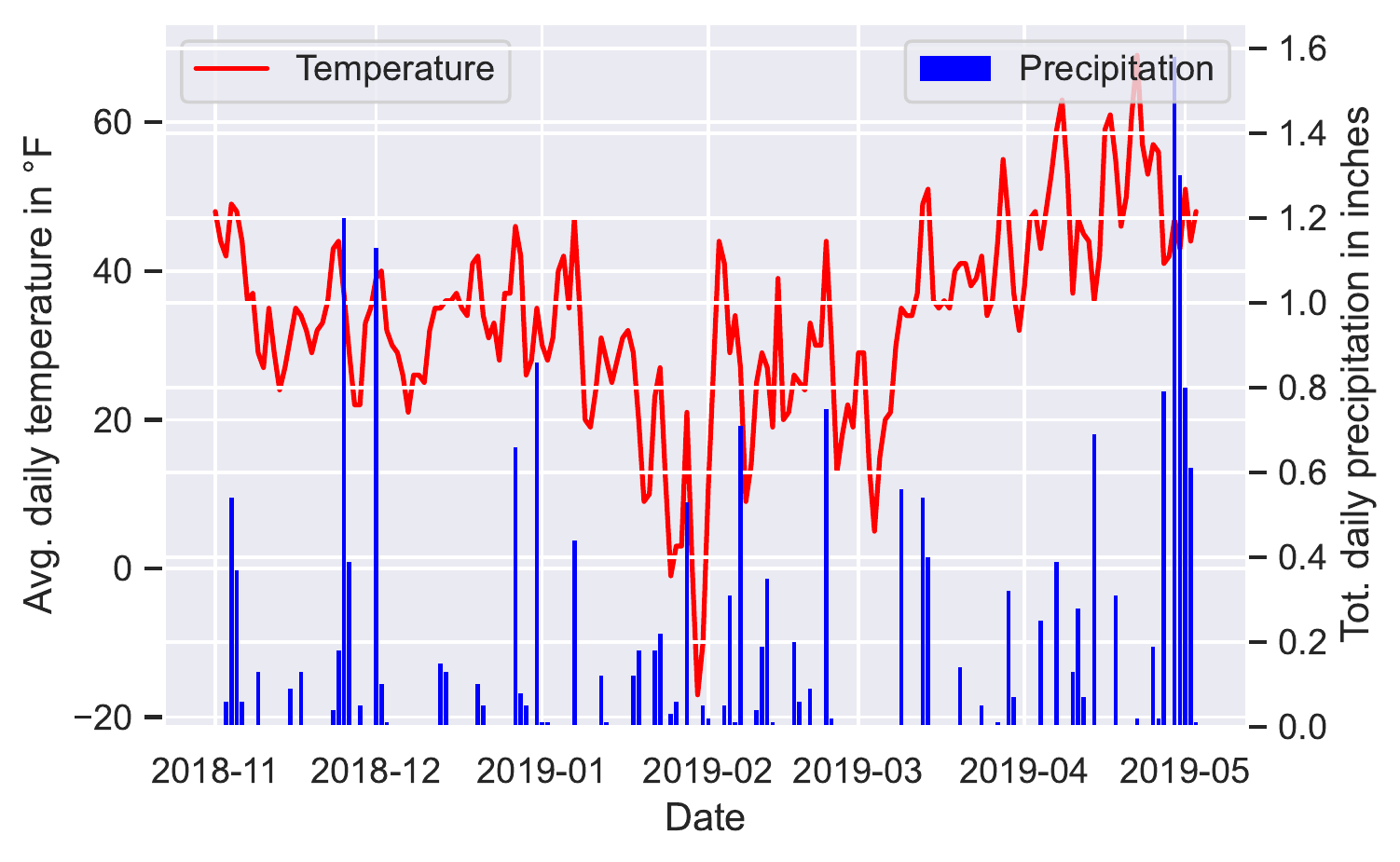}
\caption{Average daily temperature and total daily precipitation at O'Hare International Airport during the observation period} \label{fig:weather}
\end{figure}

Finally, weather information is taken from daily meteorological summaries for O'Hare International Airport provided by \citet{ncei2021daily}. 
The average daily temperature and total daily precipitation during the observation period are shown in Figure~\ref{fig:weather}.

\subsection{Imputation of mode attributes}

Lastly, we impute the attributes of the mode choice alternatives.
Driving times and distances, transit connections as well as walking and bicycling travel times are obtained from the HERE Routing Application Programming Interface\footnote{\url{https://developer.here.com/products/routing}}.

\begin{figure}[H]
\centering
\includegraphics[width = 0.6 \textwidth]{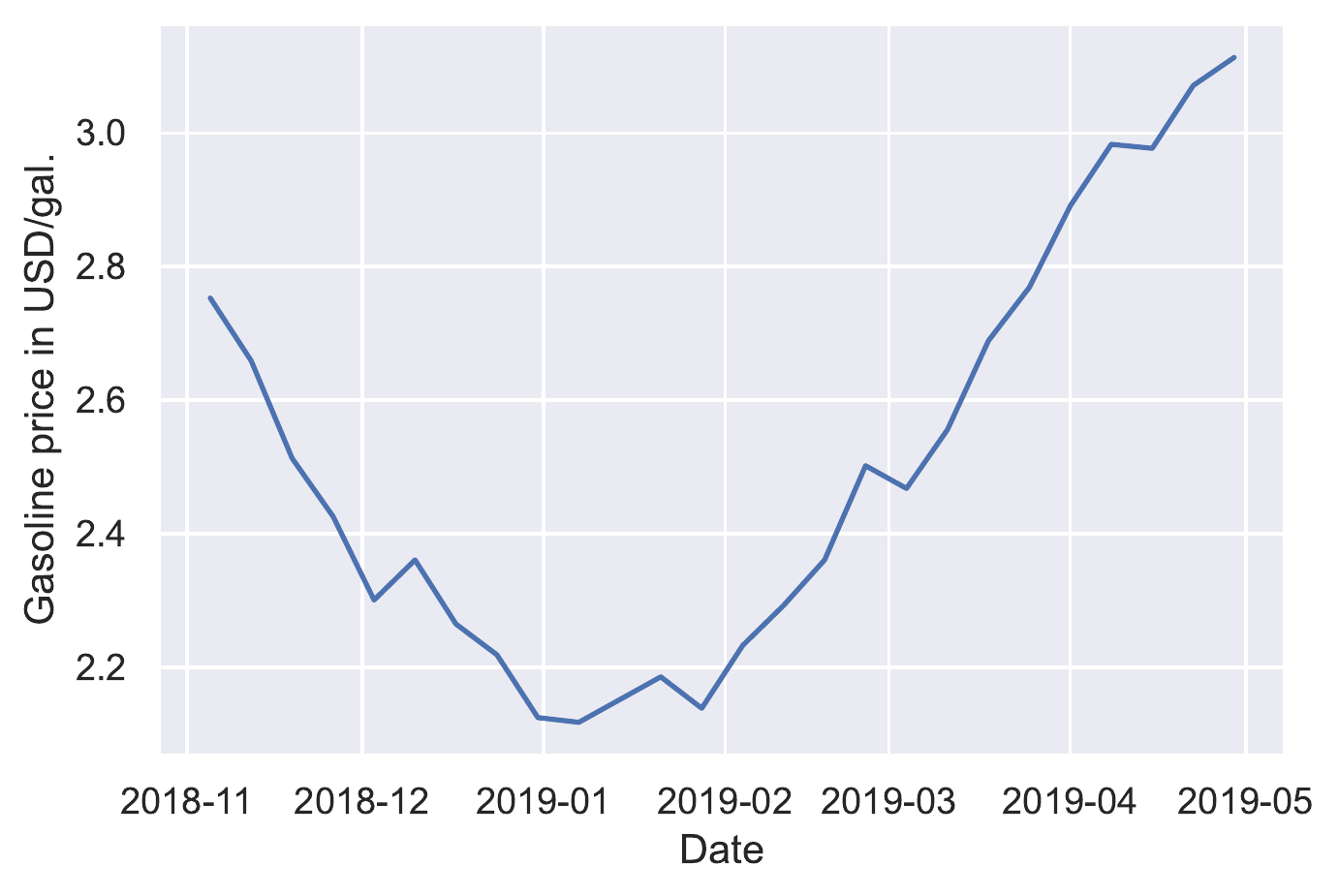}
\caption{Gasoline price during the observation period} \label{fig:gas}
\end{figure}

For the calculation of the cost of the driving alternative, we consider two variable cost components, a vehicle running cost component of 0.20 \$/mile and a fuel cost component. To compute the latter, we assume a fuel economy of 20 miles per gallon. 
Weekly average retail gasoline prices are sourced from  \citet{eia2021chicago}. 
Figure~\ref{fig:gas} shows the evolution of the unit price of gasoline during the observation period. 
We approximate transit fares using agency-specific revenue information provided in the 2019 National Transit Database \citep{fta2021ntd}. 
Based on this information, we assume a fare of 0.50 \$/mile for bus and a fare of 0.30 \$/mile for rail and metro.

Since passenger wait times for ride-sourcing and taxi are not observed, we assume a fixed wait time of two minutes for solo ride-sourcing, pooled ride-sourcing and taxi. These waiting times are included in the total travel times of these alternatives. 
To account for possible detours for picking up other passenger during pooled ride-sourcing trips, we add a ten percent travel time penalty to the driving time of pooled ride-sourcing. 

We use random forests to impute solo and pooled ride-sourcing fares. 
The random forest models take into account 
lagged fare information (i.e. the 25\textsuperscript{th}, 50\textsuperscript{th} and 95\textsuperscript{th} percentiles of the per kilometre fare in the whole network during the 30-minute time period preceding the start time of the trip),
trip attributes (driving distance and time, start time, the day of the week),
atmospheric conditions (average daily temperature, precipitation) 
as well as various attributes of the origin and destination census tracts.  
Taxi fares are calculated using official fare information \citep{chicago2020taxi}.

Table~\ref{tab:summary} provides a summary of the attributes of the chosen alternatives. 

\begin{table}[H]
\centering
\footnotesize
\input{table_summary}
\caption{Summary of attributes of chosen alternatives} \label{tab:summary}
\end{table}

\section{Econometric model} \label{sec:model}

Constructing the estimation sample from two revealed preference data sources creates two challenges in the development of a discrete choice model. 
First, a sampling correction is needed to account for the enrichment of the household travel survey data with ride-sourcing and taxi trip records. 
Second, the constructed revealed preference mode choice dataset is likely to exhibit endogeneity biases, because the demand for the ride-sourcing options and the price of the ride-sourcing options are simultaneously influenced by supply-side constraints and surge pricing mechanisms.
In this section, we present a MEV-based discrete choice model with sampling and endogeneity corrections to address these challenges. 
First, we develop the sampling correction (Section~\ref{sec:sampling}). 
Here, we consider a conditional maximum likelihood estimator due to its superior efficiency properties. 
Then, we describe the endogeneity correction (Section~\ref{sec:control}). 
Here, we select the control function approach due to its simplicity. 
The reader is directed \citet{bierlaire2020sampling} for a recent review of sampling correction approaches in discrete choice analysis and to \citet{mcfadden1999chapter} for an earlier synthesis of the topic. 
\citet{guevara2015critical} provides a review of endogeneity correction approaches.

\subsection{Discrete choice analysis under non-random sampling} \label{sec:sampling}

We consider a sample of $N$ individuals indexed by $n = 1, \ldots, N$.
Every individual $n$ is observed to choose an alternative $y_n$ from the set $\mathcal{M} = \{1, \ldots, J \}$. 
We stipulate a parametric model which generates the probability that individual $n$ chooses alternative $j \in \mathcal{M}$ given explanatory variables $\boldsymbol{x}_n$ with density $\mu(\boldsymbol{x}_n)$ and the unknown parameter $\boldsymbol{\theta}$:
\begin{equation} \label{eq:choice_model}
P(j \vert \boldsymbol{x}_n ; \boldsymbol{\theta}).
\end{equation}
Random utility theory \citep{mcfadden1981econometric} posits that a rational decision-maker $n$ chooses the option $y_n$ with the highest utility from $\mathcal{M}$, i.e.
\begin{equation}
U_{ny_n} > U_{j} \; \forall \; j \in \mathcal{M} \setminus \{y_n\},
\end{equation}
whereby 
\begin{equation}
U_{nj} = V_{nj}(\boldsymbol{x}_{nj}, \boldsymbol{\theta}) + \varepsilon_{nj}
\end{equation}
denotes the utility of alternative $j \in \mathcal{M}$. The utility $U_{nj}$ is decomposed into a deterministic aspect $V_{nj}(\boldsymbol{x}_{nj}, \boldsymbol{\theta})$ and a stochastic aspect $\varepsilon_{nj}$, which is unknown to the analyst.

We further suppose that the sample consists of $S$ subsamples indexed by $s = 1, \ldots, S$. 
Each subsample $s$ is characterised by a sampling protocol involving endogenous and exogenous stratification. 
Under an exogenous sampling protocol, the analyst selects observations based on exogenous variables $\boldsymbol{x}$.\footnote{Note that the exogenous variables $\boldsymbol{x}$ in the exogenous sampling protocol need not be the same as the explanatory variables $\boldsymbol{x}$ in the choice model.}
Under an endogenous sampling protocol, the analyst selects observations based on realised choices. 

To develop a choice model considering both exogenous and endogenous stratification, we let 
\begin{equation}
R_s (j, \boldsymbol{x}_n) = P(s \vert j, \boldsymbol{x}_n)
\end{equation}
denote the probability that a population member with configuration $\{j, \boldsymbol{x}_n \}$ qualifies for the subpopulation from which subsample $s$ is drawn. 
Consequently, the joint probability of observing a case with configuration $\{j, \boldsymbol{x}_n \}$ that qualifies for subsample $s$ is 
$R_s (j, \boldsymbol{x}_n) P(j \vert \boldsymbol{x}_n ; \boldsymbol{\theta}) \mu(\boldsymbol{x}_n)$,
and the following marginal probability gives the population share $Q_s$ of the subpopulation from which subsample $s$ is recruited:
\begin{equation}
Q_s = \sum_x \sum_j R_s (j, \boldsymbol{x}_n) P(j \vert \boldsymbol{x}_n ; \boldsymbol{\theta}) \mu(\boldsymbol{x}_n).
\end{equation}
Then, by Bayes' rule, the probability of observing a case with configuration $\{j, \boldsymbol{x}_n \}$ conditional on membership in subsample $s$ is given by
\begin{equation} \label{eq:prob_j_x}
P(j, \boldsymbol{x}_n \vert s; \boldsymbol{\theta}) 
= \frac{R_s (j, \boldsymbol{x}_n) P(j \vert \boldsymbol{x}_n ; \boldsymbol{\theta}) \mu(\boldsymbol{x}_n)} {Q_s}.
\end{equation}

Now, the likelihood of configuration $\{j, \boldsymbol{x}_n \}$ unconditional on subsample membership is
\begin{equation}
f(j, \boldsymbol{x}_n \vert \boldsymbol{\theta}) 
= \sum_{s=1}^{S} P(j, \boldsymbol{x}_n \vert s; \boldsymbol{\theta}) H_s,
\end{equation}
where $H_s = \frac{N_s}{N}$ is the share of subsample $s$ in the total sample.
Hence, the likelihood of a case with choice $j$ given explanatory variables $\boldsymbol{x}_n$ is 
\begin{equation}
f(j \vert \boldsymbol{x}_n, \boldsymbol{\theta}) 
= \frac{f(j, \boldsymbol{x}_n \vert \boldsymbol{\theta})}{\sum_{j' \in \mathcal{M}} f(j', \boldsymbol{x}_n \vert \boldsymbol{\theta})} 
= \frac{\sum_{s=1}^{S} P(j, \boldsymbol{x}_n \vert s; \boldsymbol{\theta}) H_s}{\sum_{j' \in \mathcal{M}} \sum_{s=1}^{S} P(j', \boldsymbol{x}_n \vert s; \boldsymbol{\theta}) H_s}.
\end{equation}
Substituting (\ref{eq:prob_j_x}) for $P(j, \boldsymbol{x}_n \vert s; \boldsymbol{\theta})$ yields a likelihood which is independent of $\mu (\boldsymbol{x}_n)$. We have
\begin{equation}
f(j \vert \boldsymbol{x}_n, \boldsymbol{\theta}) 
= \frac{\sum_{s=1}^{S} R_s (j, \boldsymbol{x}_n) P(j \vert \boldsymbol{x}_n ; \boldsymbol{\theta}) H_s / Q_s}{\sum_{j' \in \mathcal{M}} \sum_{s=1}^{S} R_s (j', \boldsymbol{x}_n) P(j' \vert \boldsymbol{x}_n ; \boldsymbol{\theta}) H_s / Q_s},
\end{equation}
which simplifies to
\begin{equation} \label{eq:cond_loglik}
f(j \vert \boldsymbol{x}_n, \boldsymbol{\theta}) 
= \frac{P(j \vert \boldsymbol{x}_n ; \boldsymbol{\theta}) \alpha_{nj}}{\sum_{j' \in \mathcal{M}} P(j' \vert \boldsymbol{x}_n ; \boldsymbol{\theta}) \alpha_{nj'}}
\end{equation}
with
\begin{equation}
\alpha_{nj}
= \sum_{s=1}^{S} \frac{R_s (j, \boldsymbol{x}_n) H_s}{Q_s}.
\end{equation}

(\ref{eq:cond_loglik}) suggests a conditional maximum likelihood estimator \citep{manski1981alternative} of the form
\begin{equation} \label{eq:pseudo_ml}
\hat{\mathcal{L}} (\boldsymbol{\theta} ; \boldsymbol{y}, \boldsymbol{x} )
= \sum_{n=1}^{N} \ln \frac{P(y_n \vert \boldsymbol{x}_n ; \boldsymbol{\theta}) \alpha_{ny_n}}{\sum_{j \in \mathcal{M}} P(j \vert \boldsymbol{x}_n ; \boldsymbol{\theta}) \alpha_{nj}},
\end{equation}
whereby the term 
\begin{equation} \label{eq:pseudo_ml_term}
\frac{P(y_n \vert \boldsymbol{x}_n ; \boldsymbol{\theta}) \alpha_{ny_n}}{\sum_{j \in \mathcal{M}} P(j \vert \boldsymbol{x}_n ; \boldsymbol{\theta}) \alpha_{nj}}
\end{equation}
needs to be adapted to the stipulated parametric form of the choice model (\ref{eq:choice_model}).

In the MEV family of discrete choice models \citep{mcfadden1978modeling}, the probability of choosing alternative $j \in \mathcal{M}$ conditional on explanatory variables $\boldsymbol{x}_n$ and parameters $\boldsymbol{\theta}$ is given by
\begin{equation}
P(j \vert \boldsymbol{x}_n ; \boldsymbol{\theta})
= \frac{\Lambda_{j} (\boldsymbol{x}_n , \boldsymbol{\theta})}{\sum_{j' \in \mathcal{M}} \Lambda_{j'} (\boldsymbol{x}_n , \boldsymbol{\theta})},
\end{equation}
where
\begin{equation}
\Lambda_{nj} \left (\boldsymbol{x}_n , \boldsymbol{\theta} = \{ \boldsymbol{\beta}, \boldsymbol{\lambda} \} \right ) = 
e^{V_{nj}(\boldsymbol{x}_n, \boldsymbol{\beta}) + \ln G_{nj} (\psi_{n1}, \ldots, \psi_{nJ}; \boldsymbol{\lambda}) }
\end{equation}
with 
\begin{equation}
\psi_{nj} = e^{V_{nj}(\boldsymbol{x}_n, \boldsymbol{\beta})}
\end{equation}
and
\begin{equation}
G_{nj}(\psi_{n1}, \ldots, \psi_{nJ}; \boldsymbol{\lambda}) = \frac{\partial G}{\partial \psi_{nj}} (\psi_{n1}, \ldots, \psi_{nJ}; \boldsymbol{\lambda}).
\end{equation}
Here, $V_{nj'}$ is the deterministic aspect of utility, which depends on explanatory variables $\boldsymbol{x}_n$ and parameter $\boldsymbol{\beta}$. 
$G(\psi_{n1}, \ldots, \psi_{nJ}; \boldsymbol{\lambda})$ is a MEV generating function with parameter $\boldsymbol{\lambda}$.

To evaluate (\ref{eq:pseudo_ml_term}) under the MEV assumption, we generalise the result presented in \citet{bierlaire2008estimation} from purely choice-based samples to a wider class of enriched samples. We have 
\begin{equation}
P(y_n \vert \boldsymbol{x}_n ; \boldsymbol{\theta}) \alpha_{ny_n}
= \frac{\Lambda_{ny_n} (\boldsymbol{x}_n , \boldsymbol{\theta}) \alpha_{ny_n}}{\sum_{j \in \mathcal{M}} \Lambda_{nj} (\boldsymbol{x}_n , \boldsymbol{\theta})}
\end{equation}
and define
\begin{equation}
\Lambda_{nj} (\boldsymbol{x}_n , \boldsymbol{\theta} )
= e^{V_{nj}(\boldsymbol{x}_n, \boldsymbol{\beta}) + \ln G_{nj} (\psi_{n1}, \ldots, \psi_{nJ}; \boldsymbol{\lambda}) + \ln \alpha_{nj}}
\end{equation}

Consequently, we obtain
\begin{equation} \label{eq:pseudo_ml_mev_term}
\frac{P(y_n \vert \boldsymbol{x}_n ; \boldsymbol{\theta}) \alpha_{ny_n}}{\sum_{j \in \mathcal{M}} P(j \vert \boldsymbol{x}_n ; \boldsymbol{\theta}) \alpha_{nj}}
= \frac{e^{V_{y_n}(\boldsymbol{x}_n, \boldsymbol{\beta}) + \ln G_{y_n} (\psi_{n1}, \ldots, \psi_{nJ}; \boldsymbol{\lambda}) + \ln \alpha_{ny_n}}}{\sum_{j \in \mathcal{M}} e^{V_{j}(\boldsymbol{x}_n, \boldsymbol{\beta}) + \ln G_{nj} (\psi_{n1}, \ldots, \psi_{nJ}; \boldsymbol{\lambda}) + \ln \alpha_{nj}}}
\end{equation}
Thus, a conditional maximum likelihood estimator for the MEV family of discrete choice models under non-random sampling is given by
\begin{equation} \label{eq:pseudo_ml_mev}
\hat{\mathcal{L}} (\boldsymbol{\theta} ; \boldsymbol{y}, \boldsymbol{x} )
= \sum_{n=1}^{N} \ln \frac{e^{V_{y_n}(\boldsymbol{x}_n, \boldsymbol{\beta}) + \ln G_{y_n} (\psi_{n1}, \ldots, \psi_{nJ}; \boldsymbol{\lambda}) + \ln \alpha_{ny_n}}}{\sum_{j \in \mathcal{M}} e^{V_{nj}(\boldsymbol{x}_n, \boldsymbol{\beta}) + \ln G_{nj} (\psi_{n1}, \ldots, \psi_{nJ}; \boldsymbol{\lambda}) + \ln \alpha_{nj}}}
\end{equation}
with
$\boldsymbol{\theta} = \{ \boldsymbol{\beta}, \boldsymbol{\lambda} \}$.

\subsection{Control function correction of endogeneity} \label{sec:control}

We partition the explanatory variables $\boldsymbol{x}$ into exogenous explanatory variables $\boldsymbol{c}$ and an endogenous explanatory variable $p$ such that $\boldsymbol{x} = \{\boldsymbol{c}, p \}$. 
Then, the utility of alternative $j \in \mathcal{M}$ is 
\begin{equation} \label{eq:utility_endogenous}
U_{nj} = V(\boldsymbol{c}_{nj}, p_{nj}, \boldsymbol{\beta}) + \varepsilon_{nj}
\end{equation}
with 
\begin{equation}
p_{nj} = \boldsymbol{z}_{nj}^{\top} \boldsymbol{\gamma} + \boldsymbol{\delta}^{\top} \boldsymbol{c}_{nj} + \xi_{nj}.
\end{equation}
Here, $\boldsymbol{z}_{nj}$ denotes a set of instruments, 
$\boldsymbol{\gamma}$ and $\boldsymbol{\delta}$ are unknown parameters,
and $\xi_{nj}$ is an error term.
The error term $\xi_{nj}$ captures the influence of unobserved attributes of alternative $j$ which impact $p_{nj}$ but are not included in $\boldsymbol{z}_{nj}$ and $\boldsymbol{c}_{nj}$.
The instruments $\boldsymbol{z}_{nj}$ and exogenous explanatory variables $\boldsymbol{c}_{nj}$ are independent of the stochastic aspect of utility $\varepsilon_{nj}$ and the stochastic disturbance $\xi_{nj}$.
Yet, the endogenous explanatory variable $p_{nj}$ is correlated with $\varepsilon_{nj}$, i.e. $\text{Cov}(p_{nj}, \varepsilon_{nj}) \neq 0$ and thus $\text{Cov}(\xi_{nj}, \varepsilon_{nj}) \neq 0$. 
Ignoring the endogeneity of $p_{nj}$ in the estimation of the choice model parameters $\boldsymbol{\theta}$ leads to inconsistent parameter estimates \citep{train2009discrete}.

The control function correction of endogeneity \citep{petrin2010control} consists of constructing a control variable which, when included into the utility specification, absorbs the aspect of $\varepsilon_{nj}$ that is correlated with $p_{nj}$. 
The utility error is decomposed as
\begin{equation} \label{eq:cf_linear}
\varepsilon_{nj} = C(\xi_{nj}, \phi) + \tilde{\varepsilon}_{nj},
\end{equation}
where $C(\xi_{nj}, \phi)$ is the control function with parameter $\phi$. 
$\tilde{\varepsilon}_{nj}$ is the residual error, which remains after conditioning out the aspect of $\varepsilon_{nj}$ that is correlated with $p_{nj}$.
The simplest specification of the control function is 
\begin{equation}
C(\xi_{nj}, \phi) = \phi \xi_{nj},
\end{equation}
whereby $\phi$ is an unknown scalar parameter. Then, the utility (\ref{eq:utility_endogenous}) writes
\begin{equation} 
U_{nj} = V(\boldsymbol{c}_{nj}, p_{nj}, \boldsymbol{\beta}) + \phi \xi_{nj} + \tilde{\varepsilon}_{nj}.
\end{equation}

A choice model with a control function correction of endogeneity is estimated in two stages.
First, the endogenous variable $\boldsymbol{p}_{j}$ is regressed on the instruments $\boldsymbol{z}_{j}$ and exogenous variables $\boldsymbol{c}_{nj}$. The residuals $\widehat{\xi}_{nj}$ from this regression are used to calculate the control function. 
In the second stage, the choice model is estimated, with the control function being included in the utility equation. 

\section{Model specification} \label{sec:specification}

\subsection{Second stage: Discrete choice model}

In our analysis of the mode choice dataset described in Section~\ref{sec:data}, the second stage of the two-stage model introduced in the previous section is specified as a multinomial logit model. 
We also explored various nested and cross-nested logit model specifications using the conditional maximum likelihood estimator exhibited in (\ref{eq:pseudo_ml_mev}), but no meaningful nesting structures emerged. 

The multinomial logit model assumes a specification of the random utility of the following form:
We let
\begin{align}
U_{nj} & = 
V_{nj}
+ \varepsilon_{nj} 
& & \forall \; j \in \{\text{car}, \text{transit}, \text{bike}, \text{walk}, \text{taxi} \} \label{eq:util} \\
U_{nj} & = 
V_{nj}
+ \phi_{j} \widehat{\xi}_{nj}
+ \tilde{\varepsilon}_{nj} 
& & \forall \; j \in \{\text{solo ride-sourcing}, \text{pooled ride-sourcing} \} \label{eq:util_control}
\end{align}
with 
\begin{equation}
V_{nj} =
\boldsymbol{x}_{nj}^{\text{(alt.-spec.)}} \boldsymbol{\beta}^{\text{(alt.-spec.)}}
+ \boldsymbol{x}_{nj}^{\text{(trip-spec.)}} \boldsymbol{\beta}_{j}^{\text{(trip-spec.)}} 
\quad \forall \; j \in \mathcal{M}.
\end{equation}
Here, $\boldsymbol{x}_{nj}^{\text{(alt.-spec.)}}$ and $\boldsymbol{x}_{nj}^{\text{(trip-spec.)}}$ denote alternative- and trip-specific attributes, respectively. 
The corresponding parameters are denoted by $\boldsymbol{\beta}^{\text{(alt.-spec.)}}$ and $\boldsymbol{\beta}_{j}^{\text{(trip-spec.)}}$, respectively. 
Whereas alternative-specific attributes vary across alternatives and trips (e.g. travel time, travel cost etc.), trip-specific attributes only vary across trips (e.g. census tract attributes at the origin and destination).
The parameters $\boldsymbol{\beta}_{j}^{\text{(trip-spec.)}}$ pertaining to trip-specific attributes are necessarily alternative-specific. 
For identification, we fix $\boldsymbol{\beta}_{\text{car}}^{\text{(trip-spec.)}}$ to zero.

We also incorporate alternative-specific departure time preferences in the utility specification.
Following earlier studies on air-travel itinerary choices \citep{koppelman2008schedule, lurkin2017accounting, wen2020incorporating}, we consider continuous representations of departure time preferences using a weighted sum of sine and cosine functions.
The specification has the following form:
\begin{equation}
\begin{split}
V_{nj} = \ldots 
& + \beta_{j,1} \sin \left ( \frac{2 \pi t_{n}}{1440} \right )
+ \beta_{j,2} \sin \left ( \frac{4 \pi t_{n}}{1440} \right )
+ \beta_{j,3} \sin \left ( \frac{6 \pi t_{n}}{1440} \right ) \\
& + \beta_{j,4} \cos \left ( \frac{2 \pi t_{n}}{1440} \right )
+ \beta_{j,5} \cos \left ( \frac{4 \pi t_{n}}{1440} \right )
+ \beta_{j,6} \cos \left ( \frac{6 \pi t_{n}}{1440} \right )
\end{split}
\end{equation}
Here, $\beta_{j,1}, \ldots, \beta_{j,6}$ are unknown parameters. $t_{n}$ is the observed departure time of trip $n$ in minutes past midnight, 1440 is the total number of minutes in a day.
Compared to discrete representations, continuous representations of departure time preferences produce more realistic demand predictions due to their smoothing properties \citep{koppelman2008schedule, lurkin2017accounting}.  

In accordance with the model formulation put forward in the previous section, $\phi_{j}\widehat{\xi}_{nj}$ in (\ref{eq:util_control}) is the control function, and $\phi_{j}$ is the corresponding coefficient. 
$\varepsilon_{nj}, \tilde{\varepsilon}_{nj}$ are error terms, which are assumed to be independent and identically distributed according to $\text{Gumbel}(0,1)$ across $n, t$. $\tilde{\varepsilon}_{nj}$ is the residual utility error that remains after conditioning on the aspect of $\varepsilon_{nj}$ that is correlated with the endogenous variable.

\subsection{First stage: Control function}

We hypothesise that the prices of solo and pooled ride-sourcing are endogenous, because the demand for the ride-sourcing options and their prices are co-determined by supply-side constraints and surge pricing mechanisms. 
We employ the control function approach described in Section~\ref{sec:control} to correct for this price endogeneity.
To form the control function, we must find suitable instruments that are i) correlated with the endogenous variable (i.e. price) and ii) not correlated with the error term of the demand equation. 
\citet{nevo2000practitioner} distinguishes three types of demand-side instrument, namely 
i) cost shifters,
ii) non-price attributes of other alternatives---also referred to as BLP-type instruments \citep{berry1995automobile}---and 
iii) prices of the same alternative in other markets---also referred to as Hausman-type instruments \citep{hausman1994competitive, hausman1996valuation}. 

In this work, we consider cost-shifters and non-price attributes of other alternatives as demand-side instruments for the prices of solo and pooled ride-sourcing.
First, we include driving cost as a cost-shifting instruments, whereby driving cost is calculated as the driving distance in miles times the retail price of gasoline in USD per gallon divided by an assumed fuel economy of 20 miles per gallon.
Second, we consider the aggregate frequency of transit service per capita in the origin census tract as BLP-type instruments.

\subsection{Estimation practicalities}

The multinomial logit models are estimated using the conditional maximum likelihood estimator given in (\ref{eq:pseudo_ml_mev}).
Note that the presented estimator is fully general in that it can account for both endogenous and exogenous stratification. 
In the current application, the sampling protocol is purely choice-based and does not involve stratification by an exogenous variable. 
Thus, we have $\ln \alpha_{nj} = \ln \alpha_{j} \; \forall \; n \in \{1, \ldots, N \}$. 
$\alpha_{j}$ is given by the sampling protocol defined in Table~\ref{tab:shares}.
Specifically, we have $\alpha_{j} = H_{j} / Q_{j}$, whereby $H_{j}$ is the share of observations in the sample choosing alternative $j$ (see column ``Estimation sample---Share'' in Table~\ref{tab:shares}), and
$Q_{j}$ is the share of the population choosing alternative $j$ (see column ``Population---Share'' in Table~\ref{tab:shares}). 

We implement the conditional maximum likelihood estimator for the multinomial logit models using PandasBiogeme \citep{bierlaire2018pandasbiogeme}.
The standard errors of the parameters of the discrete choice model with a control function correction are bootstrapped using 100 resamples. 
The first-stage regressions of the two-stage model are estimated using ordinary least squares.

\section{Results} \label{sec:results}

In Table~\ref{tab:results_second_stage}, we provide the estimation results of an uncorrected model (without control function correction of endogeneity but with sampling correction) and a corrected model (with endogeneity and sampling corrections). 
The first-stage results of the two-stage model are presented in Table~\ref{tab:results_first_stage}.
Summary statistics for the first-stage regressions, namely $F$-statistics, the associated $p$-values and the coefficients of determination $R^2$ are given in Table~\ref{tab:results_first_stage_sum}.
Summary statistics for the second-stage models are given in Table~\ref{tab:results_second_stage_sum}.

First, we test for the presence of endogeneity.
Under the null hypothesis that the prices of the ride-sourcing alternatives are exogenous, the second-stage coefficients $\phi_{\text{solo ride-sourcing}}$ and $\phi_{\text{pooled ride-sourcing}}$ on the first-stage residuals are zero.
Note that the considered uncorrected model is nested within the considered corrected model, since the uncorrected model can be obtained from the corrected model by setting $\phi_{\text{solo ride-sourcing}}$ and $\phi_{\text{pooled ride-sourcing}}$ equal to zero.
Under the null hypothesis that the prices of the ride-sourcing alternatives are exogenous, the restrictions imposed by the uncorrected model are supported by the observed data.
Table~\ref{tab:results_second_stage_sum} shows that the log-likelihood of the uncorrected model is $-75478.51$, while the log-likelihood of the corrected model is $-75405.76$.
A log-likelihood ratio test indicates that the improvement in fit offered by the corrected model is statistically significant ($\tilde{\chi}^{2} = 145.51$, $\text{df} = 2$, $p < 0.001$). 
Thus, we reject the constraints of the uncorrected model and conclude that the prices of the ride-sourcing alternatives are endogenous.

In the corrected model, the estimates of the parameters pertaining to mode attributes have the expected signs and are significantly different from zero. 
More precisely, the mode-specific travel time parameters are all negative and statistically significant. 
As expected, a larger number of transfers appears to decrease the propensity of choosing public transit, and a higher hourly park rate at the destination appears to decrease the propensity of choosing car.
Strikingly, the estimate of $\beta_{\text{cost, car}}$ is not statistically significant in the uncorrected model.

In Table~\ref{tab:elas_own}, we compare the weighted direct aggregate arc elasticities with respect to travel cost and time of the two models.
It can be seen that in the corrected model, the demand for taxi as well as for solo and pooled ride-sourcing is substantially more elastic with respect to price than in the uncorrected model.
For example, the estimated direct aggregate arc elasticity with respect to the cost of pooled ride-sourcing is $-0.262$ in the uncorrected model and $-0.923$ in the corrected model. 
The corrected model further reveals that the demand for the considered mode choice alternatives is elastic with respect to travel time. 
As expected, walking and biking exhibit the highest elasticities with respect to travel time. 
Taxi and pooled ride-sourcing are more elastic with respect to travel time than solo ride-sourcing. 

\begin{figure}[H]
\centering
\includegraphics[width = 0.6 \textwidth]{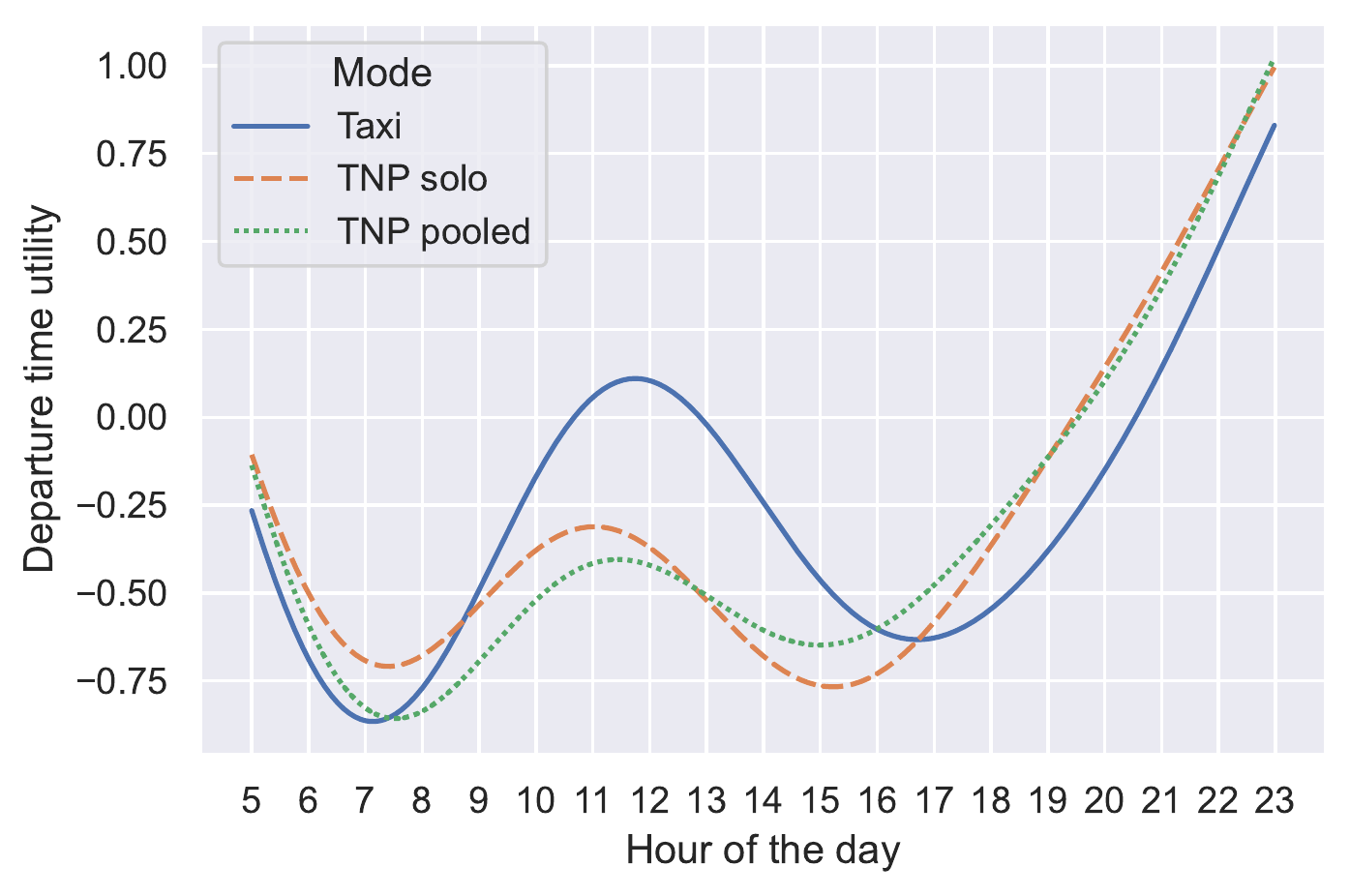}
\caption{Departure time utility} \label{fig:dep}
\end{figure}

Figure~\ref{fig:dep} visualises the estimated continuous departure time preferences in the corrected model.
While there are minor differences in departure time preferences across the three modes in the morning, afternoon and evening hours, solo and pooled ride-sourcing appear to be comparatively less likely to be chosen mid-day.

We also observe that weather conditions affect the demand for taxi and ride-sourcing.
Our utility specification includes both main and interaction effects of the average daily temperature and the daily precipitation amount.
To facilitate the interpretation of the effects, we standardised the former and kept the latter on its original scale. 
The estimates of these effects are statistically significant for taxi as well as solo and pooled ride-sourcing. 
Since the estimated effects have the same signs and are in the same order of magnitude, the same interpretation applies to the estimated effects for all three modes. 
At the mean average daily temperature in the observation period, positive precipitation increases the demand for taxi and ride-sourcing. 
On dry days, a higher temperature leads to increased demand for taxi and ride-sourcing. 

\begin{figure}[H]
\centering
\includegraphics[width = 0.48 \textwidth]{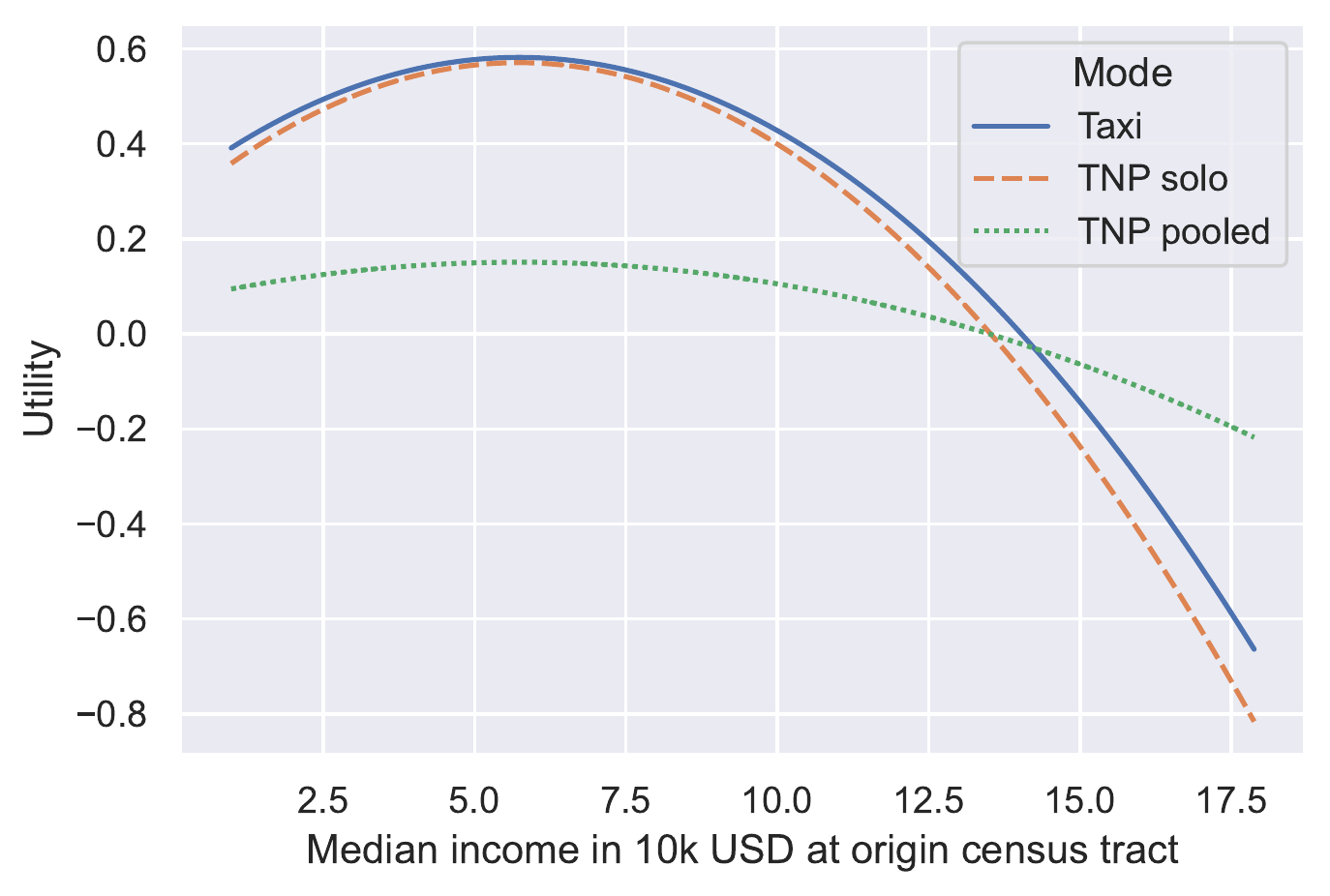}
\includegraphics[width = 0.48 \textwidth]{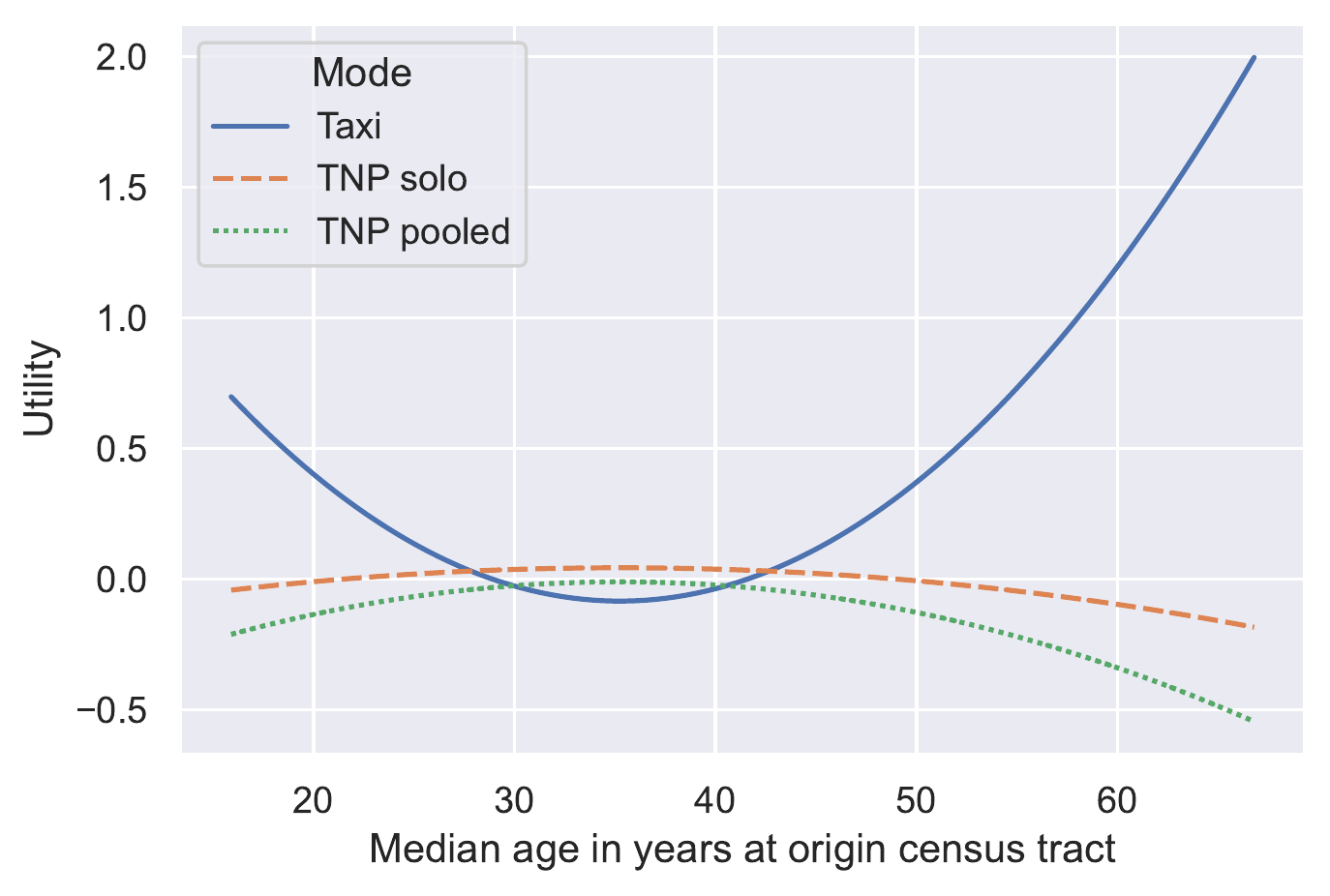}
\caption{Income and age effects in the corrected model} \label{fig:sen_income_age}
\end{figure}

The corrected model also provides insights into the influence of census tract attributes on travel demand.
Due to the inclusion of quadratic terms in the utility specification, we are able to capture non-linear income and age effects on the demand for taxi as well as solo and pooled ride-sourcing. 
These effects are visualised in Figure~\ref{fig:sen_income_age} over their respective realised ranges in the training dataset. 
For all three modes, the non-linear income effects are concave down, whereby the curvature is more pronounced for taxi and solo ride-sourcing than for pooled ride-sourcing. 
The demand for pooled ride-sourcing appears to be less sensitive to income compared to the demand for taxi and solo ride-sourcing.
Increasing income initially has a positive effect on the demand for taxi and solo ride-sourcing, but the effect of income becomes negative for median annual household incomes above USD 140,000. 
This suggests that the demand for taxi and solo ride-sourcing is comparatively lower in census tracts with high household incomes. 

Next, we consider the estimated age effects in the corrected model.
For solo and pooled ride-sourcing, the age effects are concave down, while they are concave up for taxi. 
The curvature of the age effect on demand for taxi is substantially more pronounced than for the other two modes. 
In comparison to the age effect for taxi, solo and pooled ride-sourcing do not appear sensitive to age. 
The effect of age on taxi demand increases sharply for median ages above 35 years, 
which suggests that taxi demand is comparatively higher in census tracts with older residents. 

Various land use and built environment characteristics also influence the demand demand for taxi and ride-sourcing. 
For example, a higher residential density increases the propensities of choosing taxi and solo ride-sourcing. 
A higher employment density increases the propensity of choosing taxi but decreases the propensity of choosing ride-sourcing.
A higher land use diversity decreases the propensities of choosing taxi and ride-sourcing. 
A denser network of pedestrian-oriented links decreases the propensities of choosing taxi and ride-sourcing.
However, a higher intersection density at the trip origin increases the propensities of choosing taxi and ride-sourcing.

\begin{table}[H]
\centering
\notsotiny
\input{table_results_second_stage_formatted}
\caption{Second stage estimation results} \label{tab:results_second_stage}
\end{table}

\begin{table}[H]
\centering
\footnotesize
\input{table_results_first_stage}
\caption{First stage estimation results (instruments only)} \label{tab:results_first_stage}
\end{table}

\begin{table}[H]
\centering
\small
\input{table_results_first_stage_sum}

\caption{First stage estimation summary} \label{tab:results_first_stage_sum}
\end{table}

\begin{table}[H]
\centering
\small
\input{table_results_second_stage_sum}
\caption{Second stage estimation summary} \label{tab:results_second_stage_sum}
\end{table}

\begin{table}[H]
\centering
\small
\input{table_elas_own}
\caption{Aggregate direct point elasticities} \label{tab:elas_own}
\end{table}

\section{Welfare analysis} \label{sec:welfare}

We also use the corrected model to analyse the welfare implications of ride-sourcing. 
More specifically, we consider three scenarios in which we simulate welfare losses due to the removal of i) all ride-sourcing, ii) solo ride-sourcing and iii) pooled ride-sourcing services from the choice sets of observations in which the removed mode is the chosen mode.
In addition, we analyse the welfare implications of ride-sourcing taxes, inspired by a congestion tax implemented in Chicago in 2020 \citep{mcmahon2020citys}.
More specifically, we consider two taxation scenarios in which a tax is added to the fare of trips in which solo ride-sourcing is selected.
In the first scenario, we impose a fixed tax of USD 3 on solo ride-sourcing trips, and in the second scenario, we apply a variable tax of 20\% added to solo ride-sourcing fares.

For each scenario and trip, we compute compensating variations, i.e.\ the monetary compensations that offset the alteration of the choice sets. 
Since the considered model includes alternative-specific cost parameters, it is not possible to analytically compute compensating variations. 
Therefore, we adopt the simulation approach presented in \citet{mcfadden2012computing}. For completeness, we also describe the approach in Appendix~\ref{app:compensating_variations}.

\begin{figure}[H]
\centering
\includegraphics[width = 0.6 \textwidth]{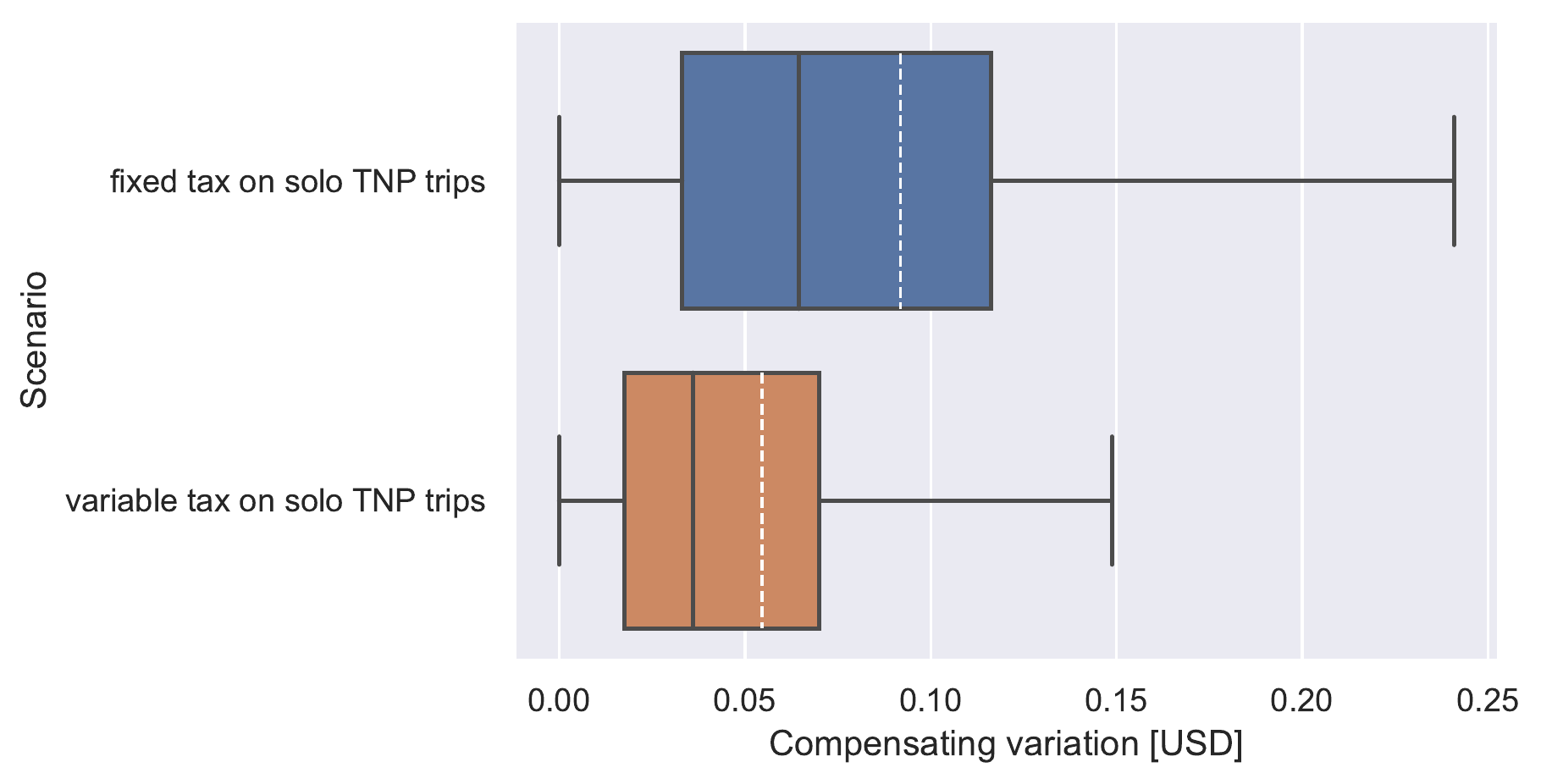} 
\caption{Box plots of compensating variations in taxation scenarios (dashed white lines indicate means)} \label{fig:box_compensating_variations_tax}
\end{figure}

In Figures~\ref{fig:box_compensating_variations_elim} and ~\ref{fig:box_compensating_variations_tax}, we show box plots of the computed compensating variations in the elimination and taxation scenarios, respectively.
Figure~\ref{fig:box_compensating_variations_elim} suggests that welfare losses are largest due to the elimination of all ride-sourcing services, closely followed by the removal of only solo ride-sourcing services, whereas welfare losses due to the removal of pooled ride-sourcing services are comparatively small.
While the mean compensating variations for the first two elimination scenarios are USD 0.44 and USD 0.35, respectively, the compensating variation in the third scenario, in which only pooled ride-sourcing services are eliminated, is USD 0.13.
As expected, welfare losses in the taxation scenarios are smaller compared to the elimination scenarios (see Figure~\ref{fig:box_compensating_variations_tax}).
This is because in the taxation scenarios alternatives are made less attractive through the introduction of a tax but are not entirely removed from choice sets.
It can be seen that welfare losses are higher in the scenarios with a fixed tax compared to the scenarios with a variable tax.
Whereas the mean compensating variation is USD 0.09 in the first taxation scenario, the mean compensating variation in the second scenario is USD 0.05.
In all scenarios, the distributions of the compensating variations exhibit a considerable spread and are right-tailed. 
For example, in the first elimination scenario, the interquartile range of the compensating variations is USD 0.41, and in the first taxation scenario the interquartile range is USD 0.12.
In all scenarios, the mean is larger than the median.
These results suggest that the distribution of ride-sourcing benefits is highly heterogeneous.
Overall, the compensating variations appear small. However, this can be explained by the fact that the ride-sourcing alternatives have comparatively small probabilities of being selected compared to other alternatives, such as car and public transit. 

\begin{figure}[H]
\centering
\includegraphics[width = 0.3 \textwidth]{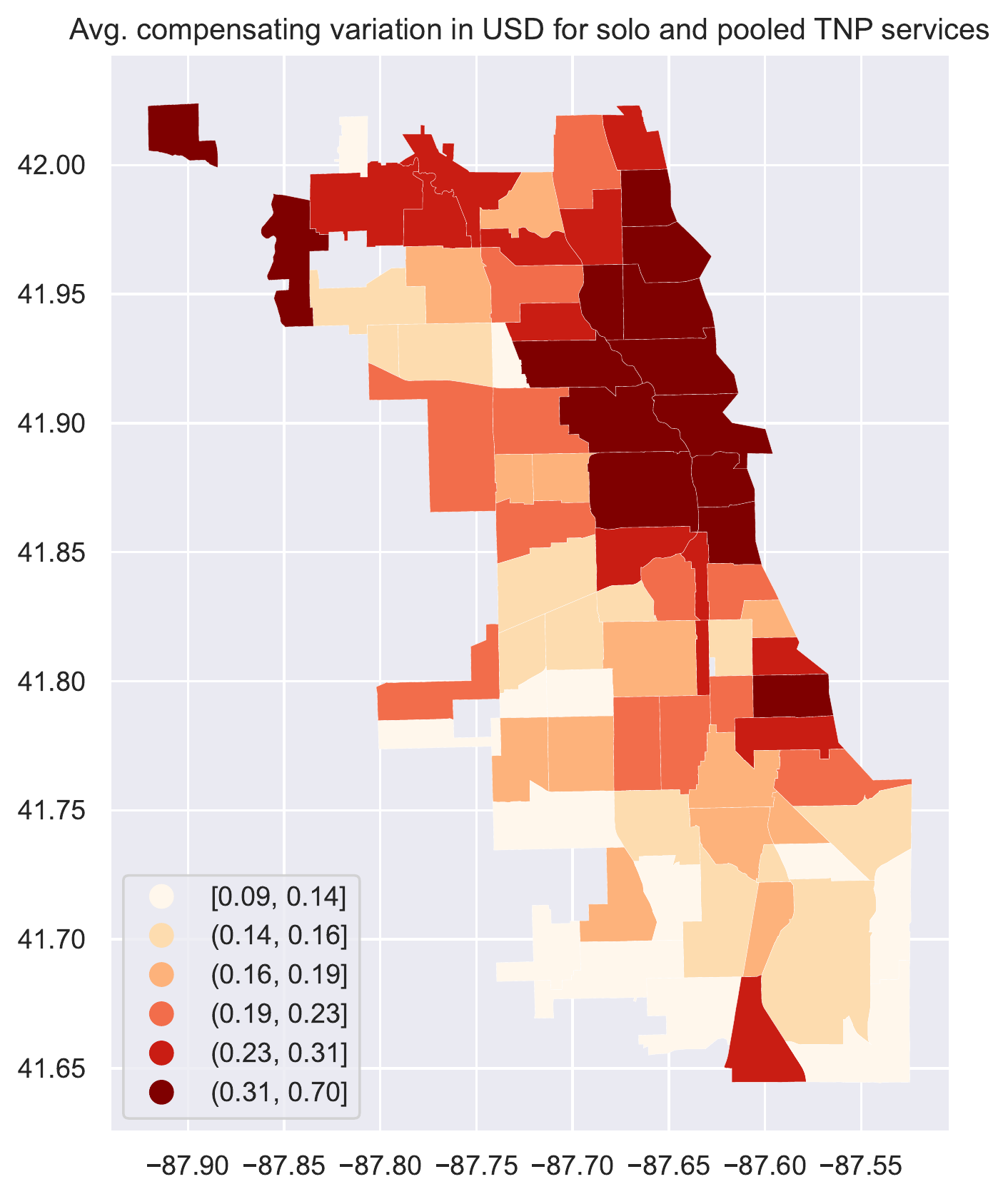} 
\includegraphics[width = 0.3 \textwidth]{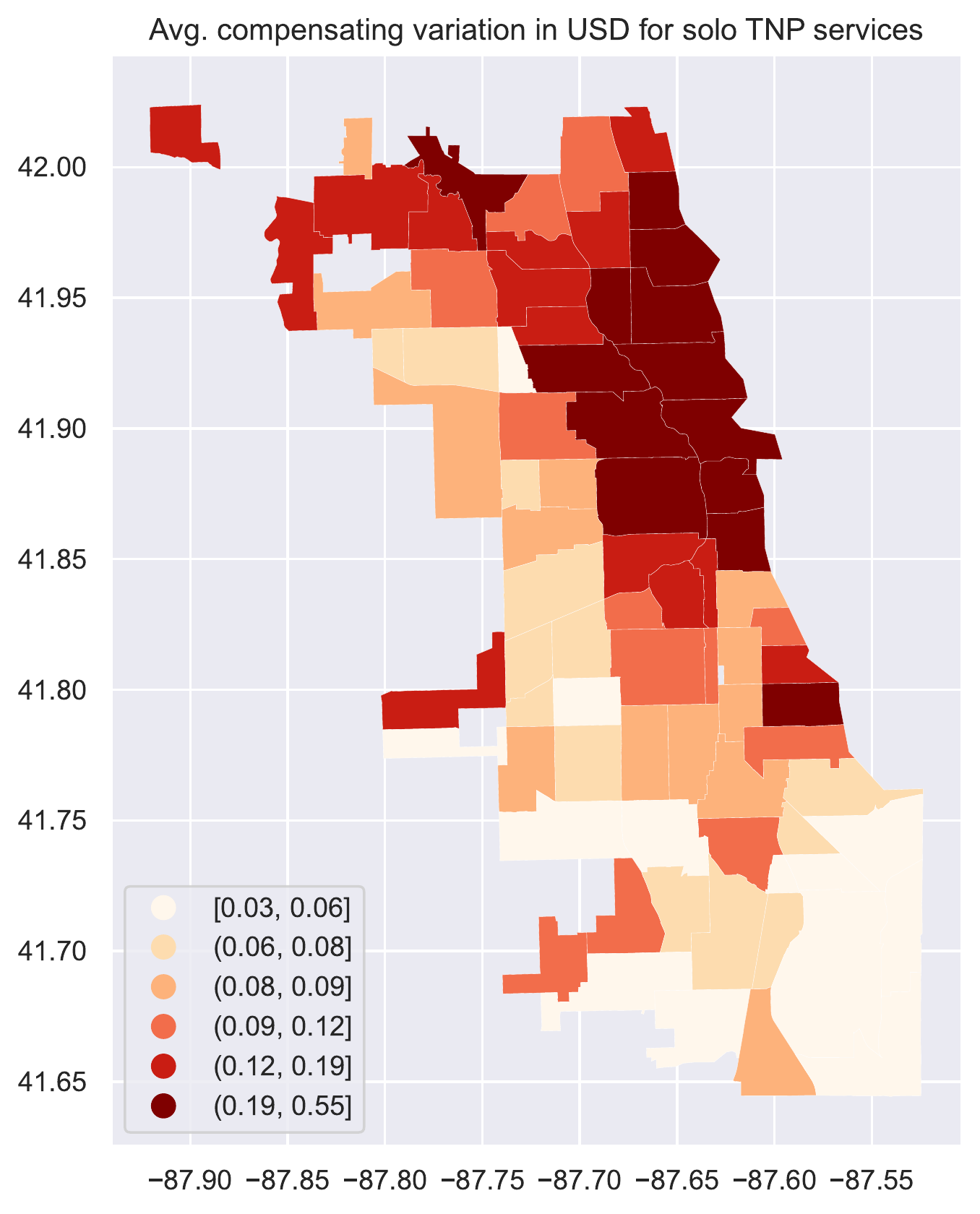}
\includegraphics[width = 0.3 \textwidth]{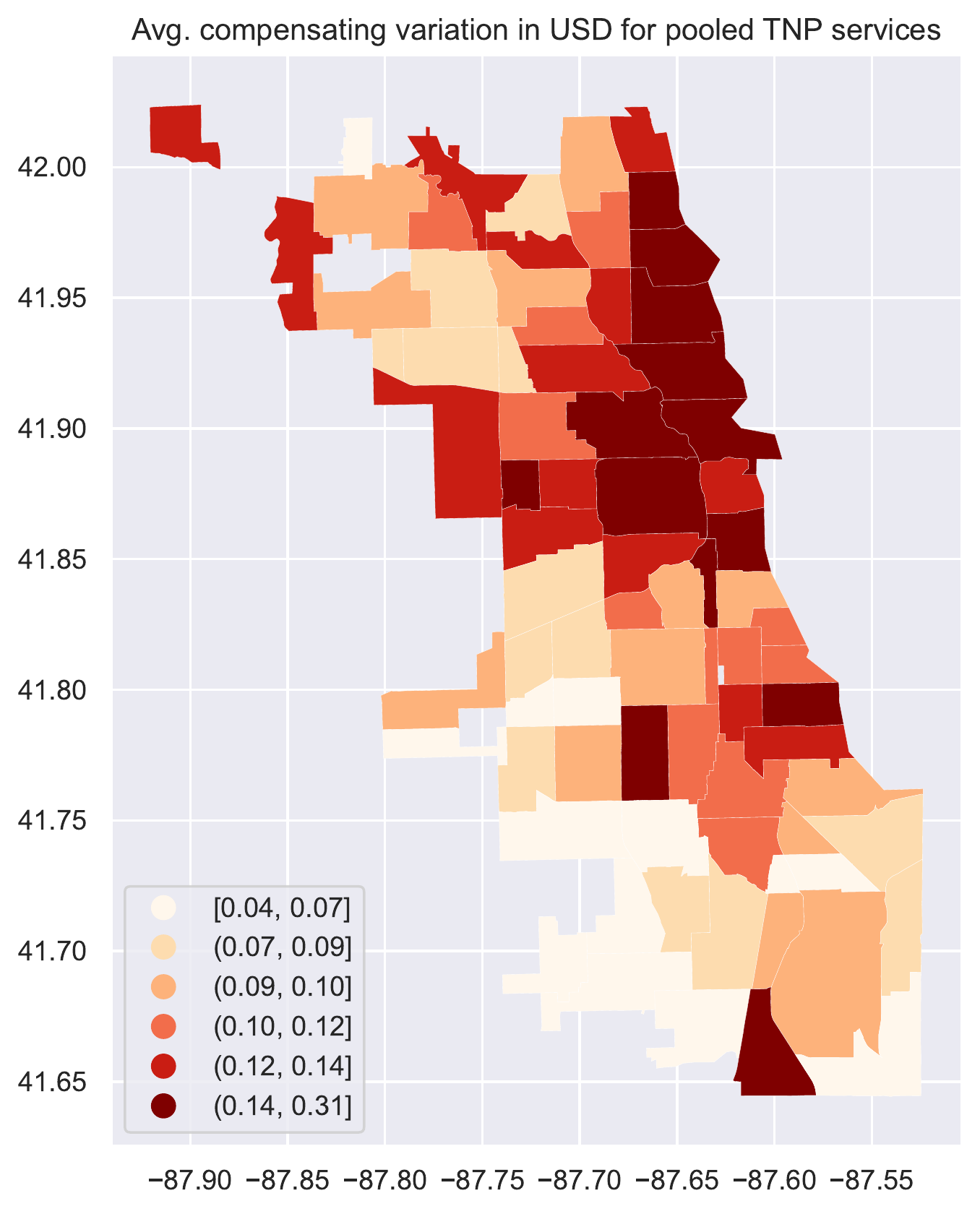}
\caption{Average compensating variations by community area in elimination scenarios} \label{fig:compensating_variations_elim}
\end{figure}

In Figure~\ref{fig:compensating_variations_elim}, we present the average compensating variations by community area for the elimination scenarios, in which services are removed from choice sets. 
The figure reveals substantial heterogeneity in the distributions of the computed compensating variations in the three considered scenarios across the community areas of the study region.
In all three scenarios, ride-sourcing benefits are valued higher in central areas. 
The average compensating variations in the community areas of the study area in the first scenario range from USD 0.09 to 0.70. 
Benefits of solo ride-sourcing are valued higher than the benefits of pooled ride-sourcing. Whereas the average compensating variations in the second scenario range from USD 0.03 to 0.55, the average compensating variations in the third scenario range from USD 0.04 to only 0.31.

\begin{figure}[H]
\centering
\includegraphics[width = 0.3 \textwidth]{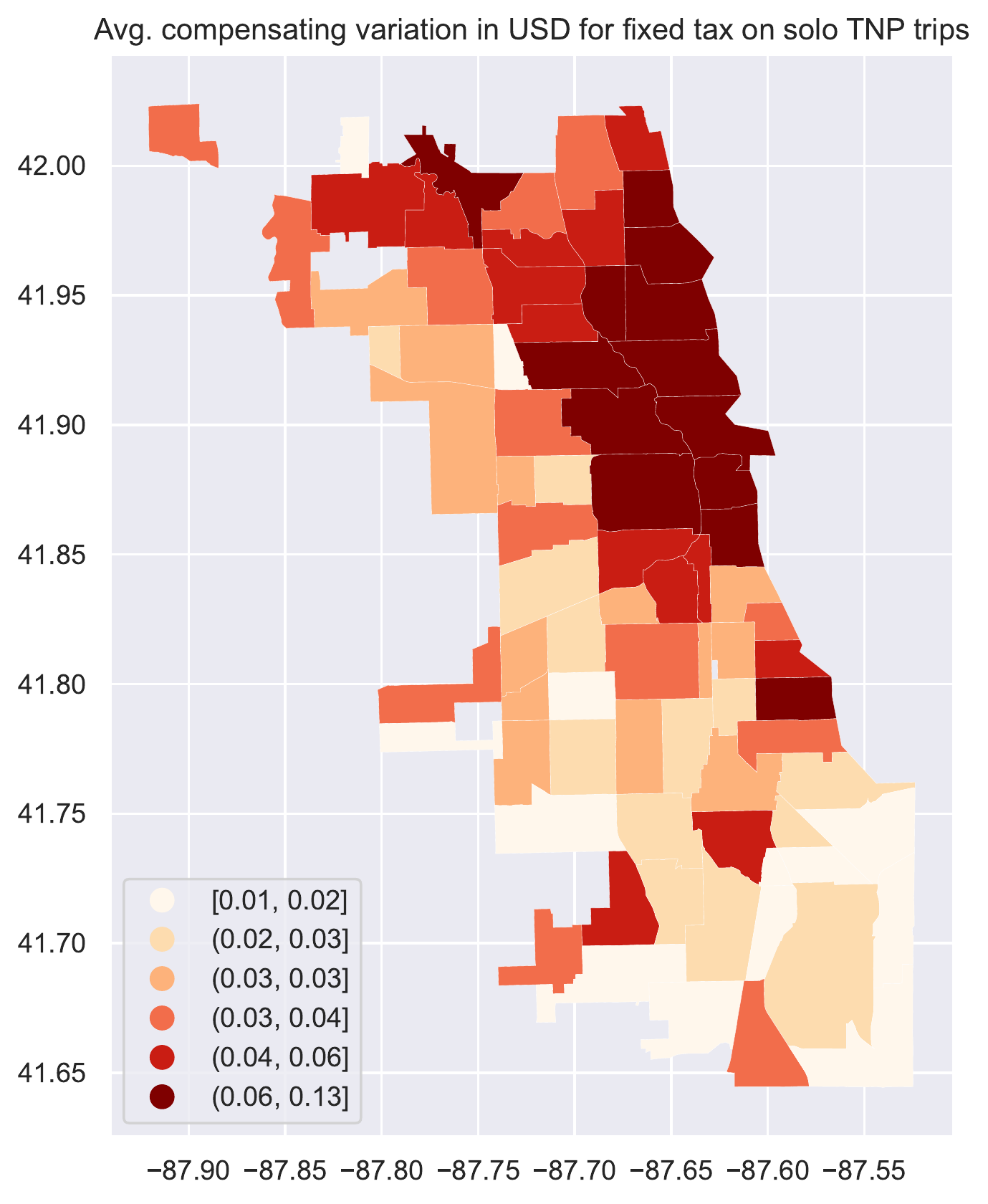} 
\includegraphics[width = 0.3 \textwidth]{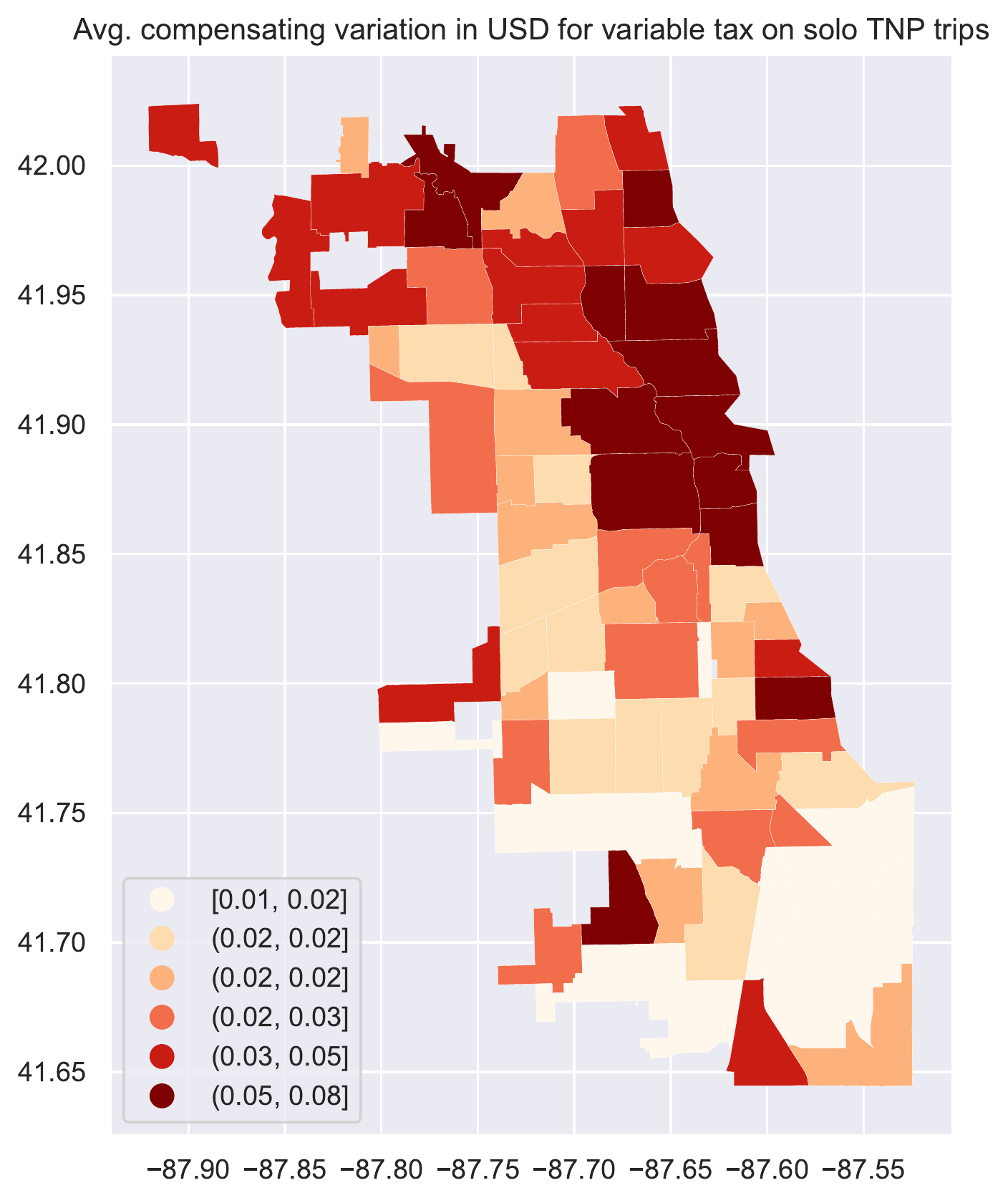}
\caption{Average compensating variations by community area in taxation scenarios} \label{fig:compensating_variations_tax}
\end{figure}

Finally, in Figure~\ref{fig:compensating_variations_tax}, we present the average compensating variations by community area for the taxation scenarios. 
The average compensating variations range from USD 0.01 to 0.13 in the fixed tax scenario and from USD 0.01 to 0.0.08 in the variable tax scenarios.
The spatial distributions of the compensating variations appear similar in both scenarios and are consistent with the spatial distributions obtained in the elimination scenarios. 

\begin{figure}[H]
\centering
\includegraphics[width = 0.6 \textwidth]{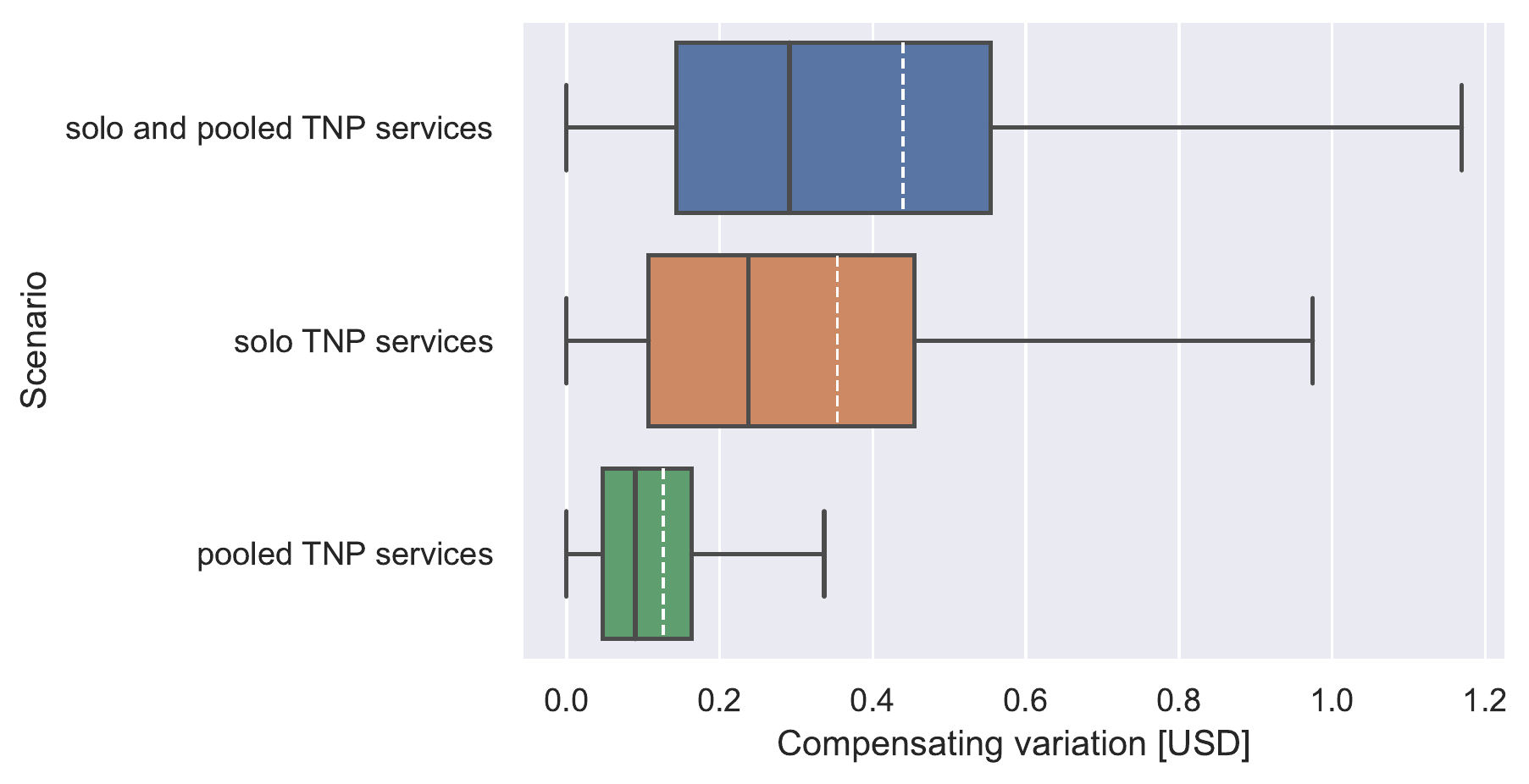} 
\caption{Box plots of compensating variations in elimination scenarios (dashed white lines indicate means)} \label{fig:box_compensating_variations_elim}
\end{figure}

\section{Conclusion} \label{sec:conclusion}

In this paper, we presented and applied an approach for estimating ride-sourcing demand at a disaggregate level from multiple data sources using DCMs. 
In sum, our research makes four contributions to the literature. First, we demonstrate how ride-sourcing demand estimation with DCMs can be performed by fusing multiple disaggregate data sources. Second, we show how traditional household travel surveys can be enriched with emerging sources of big data (i.e. trip records). Third, we highlight the importance of controlling for endogeneity biases in ride-sourcing demand estimation. Finally, we provide a methodology for incorporating emerging mobility options (such as ride- and bike-sharing etc.) into disaggregate activity-based travel demand forecasting models. 

There are several ways in which our work could be extended. First, an integrated choice and latent variable model could be adopted to accommodate flexible substitution patterns and to simultaneously estimate the two model stages. Second, the constructed mode choice dataset could be enriched with trip records providing information about other emerging transport modes such as bike-sharing. Third, the temporal structure of the data could be explicitly considered to investigate the temporal stability of the structural relationship between travel demand and the various explanatory variables. 

\section*{Author contribution statement}

\textbf{Rico Krueger:} Conceptualisation, Methodology, Software, Formal analysis, Investigation, Data curation, Writing – original draft, Visualisation.
\textbf{Michel Bierlaire:} Conceptualisation, Methodology, Writing – review \& editing.
\textbf{Prateek Bansal:} Conceptualisation, Methodology, Investigation, Writing – original draft.

\newpage
\bibliographystyle{apalike}
\bibliography{bibliography.bib}

\newpage

\begin{appendices}

\section{Overview of ride-sourcing demand analysis studies} \label{app:lit_review}

\begin{landscape}
\centering
\footnotesize
\input{table_lit_review}
\end{landscape}

\section{Computation of compensating variations} \label{app:compensating_variations}

We adopt the simulation approach presented in \citet{mcfadden2012computing} for the computation of the compensating variations, i.e.\ the monetary compensations for offsetting exogenous changes to choice sets and alternatives.
The approach is implemented as follows.
Suppose that a policy changes the choice set from $\mathcal{C}^{0}$ to $\mathcal{C}^{1}$, prices from $p_{nj}^{0}$ to $p_{nj}^{1}$ and non-cost attributes from $\boldsymbol{X}_{n}^{0}$ to $\boldsymbol{X}_{n}^{1}$.
Prior to the policy change, the random utility of alternative $j \in \mathcal{C}^{0}$ is
\begin{equation}
U_{nj}(p_{nj}^{0}, \boldsymbol{X}_{n}^{0}, \varepsilon_{nj}) = p_{nj}^{0} \beta_{p_{j}} + \left ( \boldsymbol{X}_{n}^{0} \right )^{\top} \boldsymbol{\beta}_{\boldsymbol{X}} + \varepsilon_{nj},
\end{equation}
After the policy change, the random utility of alternative $j \in \mathcal{C}^{1}$ is
\begin{equation}
U_{nj}(p_{nj}^{1} - \nu_{n}, \boldsymbol{X}_{n}^{1}, \varepsilon_{nj}) = \left ( p_{nj}^{1} - \nu_{n} \right ) \beta_{p_{j}} + \left ( \boldsymbol{X}_{n}^{1} \right )^{\top} \boldsymbol{\beta}_{\boldsymbol{X}} + \varepsilon_{nj},
\end{equation}
where $\nu_{n}$ is the compensating variation which is implicitly given by
\begin{equation}
\underset{j \in \mathcal{C}^{0}}{\operatorname{max}} \; U_{nj}(p_{nj}^{0}, \boldsymbol{X}_{n}^{0}, \varepsilon_{nj}) =
\underset{j \in \mathcal{C}^{1}}{\operatorname{max}} \; U_{nj}(p_{nj}^{1} - 
\nu_{n}, \boldsymbol{X}_{n}^{1}, \varepsilon_{nj}).
\end{equation}
We use simulation to obtain an estimate of $\widehat{\nu}_{n}$. 
To be specific, we proceed in five steps:
\begin{enumerate}
    \item Draw $\boldsymbol{\varepsilon}_{n}^{t}$ from the estimated distribution of $\boldsymbol{\varepsilon}_{n}$. Note that drawing from the estimated distribution of $\boldsymbol{\varepsilon}_{n}$ is especially easy if the considered choice model is multinomial logit. 
    \item Conditional on $\boldsymbol{\varepsilon}_{n}^{t}$, calculate $U_{nj}^{*} = \underset{j \in \mathcal{C}^{0}}{\operatorname{max}} \; U_{nj}(p_{nj}^{0}, \boldsymbol{X}_{n}^{0}, \varepsilon_{nj})$.
    \item Also conditional on $\boldsymbol{\varepsilon}_{n}^{t}$, determine the minimal compensating variation $\nu_{n}^{t}$ that satisfies $U_{nj}^{*} = \underset{j \in \mathcal{C}^{1}}{\operatorname{max}} \; U_{nj}(p_{nj}^{1} - \nu_{n}^{t}, \boldsymbol{X}_{n}^{1}, \varepsilon_{nj})$.
    \item Repeat Steps 1--3, $T$ times to find $T$ compensating variations $\nu_{n}^{t}$.
    \item Compute $\widehat{\nu}_{n} = \sum_{t=1}^{T} \nu_{n}^{t}$.
\end{enumerate}

\end{appendices}

\end{document}

%% file: table_shares.tex
\begin{tabular}{l|c|rr|rr|rr}
\toprule
{} &  \textbf{Population} & \multicolumn{2}{c|}{\textbf{My Daily Travel}} & \multicolumn{2}{c|}{\textbf{Trip records}} & \multicolumn{2}{c}{\textbf{Estimation sample}} \\
\textbf{Mode} & \textbf{Share} &    \textbf{Count} & \textbf{Share} &    \textbf{Count} & \textbf{Share} &           \textbf{Count} & \textbf{Share} \\
\midrule
Car        & 0.521 &  4781 & 0.380 &       0.000 &   0.000 &        4781 & 0.066 \\
Transit    & 0.200 &  4275 & 0.339 &       0.000 &   0.000 &        4275 & 0.059 \\
Walk       & 0.200 &  2737 & 0.217 &       0.000 &   0.000 &        2737 & 0.038 \\
Bike       & 0.050 &   334 & 0.027 &       0.000 &   0.000 &         334 & 0.005 \\
Taxi       & 0.004 &    45 & 0.004 & 20000 & 0.333 &       20045 & 0.276 \\
Solo ride-sourcing        & 0.016 &   285 & 0.023 & 20000 & 0.333 &       20285 & 0.279 \\
Pooled ride-sourcing & 0.009 &   136 & 0.011 & 20000 & 0.333 &       20136 & 0.277 \\
\midrule
Sum        & 1.000 & 12593 & 1.000 & 60000 & 1.000 &       72593 & 1.000 \\
\bottomrule
\end{tabular}

%% file: table_summary.tex
\begin{tabular}{l|rrrrrrrr}
\toprule
Alternative \\
\quad Attribute &    Count &  Mean &   Std. &  Min. &   25\% &   50\% &   75\% &    Max. \\
\midrule
Driving \\
\quad Cost [USD]            &  4781 &  1.43 &  1.36 & 0.06 &  0.49 &  0.94 &  1.89 &   9.62 \\
\quad Travel time [min]            &  4781 & 17.34 & 13.03 & 0.97 &  7.67 & 13.67 & 22.83 &  91.00 \\
\quad Avg. park rate at destination [USD/h]           &  4781 &  1.89 &  2.11 & 0.00 &  0.05 &  1.99 &  2.00 &   9.84 \\
\midrule
Transit \\
\quad Walking time [min]      &  4275 & 17.30 &  6.70 & 0.00 & 12.00 & 16.00 & 22.00 &  52.00 \\
\quad On-vehicle time [min]        &  4275 & 18.82 & 12.26 & 0.00 &  9.00 & 17.00 & 26.00 &  86.00 \\
\quad Fare [USD]      &  4275 &  1.84 &  1.37 & 0.05 &  0.77 &  1.56 &  2.57 &  10.44 \\
\quad Transfers       &  4275 &  0.72 &  0.77 & 0.00 &  0.00 &  1.00 &  1.00 &   3.00 \\
\midrule
Walking \\
\quad Travel time [min]               &  2737 & 28.89 & 37.06 & 3.88 & 13.85 & 20.37 & 28.77 & 595.62 \\
\midrule
Bike \\
\quad Travel time [min]             &   334 & 22.21 & 15.96 & 2.70 &  9.78 & 18.18 & 30.26 &  75.20 \\
\midrule
Taxi \\
\quad Travel time [min]       & 20045 & 14.20 &  7.40 & 2.82 &  9.85 & 12.23 & 16.12 &  77.53 \\
\quad Fare [USD]               & 20045 & 13.24 &  7.97 & 3.78 &  8.93 & 10.79 & 14.61 &  90.08 \\
\midrule
Solo ride-sourcing [min] \\
\quad Travel time [min]       & 20285 & 18.13 &  9.30 & 2.87 & 11.52 & 15.83 & 22.75 &  76.70 \\
\quad Fare [USD]                & 20285 &  9.19 &  5.26 & 2.50 &  5.00 &  7.50 & 12.50 & 122.50 \\
\midrule
Pooled ride-sourcing \\
\quad Travel time [min] & 20136 & 22.81 & 11.55 & 2.94 & 14.34 & 20.11 & 28.76 &  95.78 \\
\quad Fare [USD]         & 20136 &  7.10 &  4.00 & 2.50 &  5.00 &  5.00 & 10.00 &  40.00 \\
\bottomrule
\end{tabular}

%% file: table_results_second_stage_formatted.tex
\begin{tabular}{l|lrr|lrr}
\toprule
 & \multicolumn{3}{c|}{MNL uncorrected} & \multicolumn{3}{c}{MNL with control function} \\
 & Est. & Std. err. & z-stat. & Est. & Std. err. & z-stat. \\
\midrule
ASCs & & & & \\
\quad $\text{ASC}_{\text{transit}}$                                                 &            $-1.641^{***}$ &     0.053 & -31.100 &            $-1.589^{***}$ &     0.062 & -25.665 \\
\quad $\text{ASC}_{\text{walk}}$                                                    &  $\phantom{-}0.451^{***}$ &     0.081 &   5.591 &  $\phantom{-}0.476^{***}$ &     0.086 &   5.539 \\
\quad $\text{ASC}_{\text{bike}}$                                                    &            $-2.970^{***}$ &     0.108 & -27.383 &            $-2.932^{***}$ &     0.109 & -26.885 \\
\quad $\text{ASC}_{\text{taxi}}$                                                    &            $-6.424^{***}$ &     0.116 & -55.278 &            $-6.377^{***}$ &     0.114 & -55.992 \\
\quad $\text{ASC}_{\text{solo ride-sourcing}}$                                      &            $-3.823^{***}$ &     0.109 & -34.996 &            $-3.739^{***}$ &     0.100 & -37.557 \\
\quad $\text{ASC}_{\text{pooled ride-sourcing}}$                                    &            $-4.278^{***}$ &     0.109 & -39.204 &            $-3.993^{***}$ &     0.106 & -37.568 \\
\midrule
Mode attributes & & & & \\
\quad $\beta_{\text{cost, car}}$                                                    &                  $-0.015$ &     0.040 &  -0.361 &            $-0.159^{***}$ &     0.040 &  -3.995 \\
\quad $\beta_{\text{cost, transit}}$                                                &            $-0.216^{***}$ &     0.066 &  -3.299 &            $-0.289^{***}$ &     0.068 &  -4.241 \\
\quad $\beta_{\text{cost, taxi}}$                                                   &            $-0.060^{***}$ &     0.003 & -18.795 &            $-0.085^{***}$ &     0.005 & -16.250 \\
\quad $\beta_{\text{cost, solo ride-sourcing}}$                                     &            $-0.056^{***}$ &     0.005 & -10.251 &            $-0.114^{***}$ &     0.011 & -10.305 \\
\quad $\beta_{\text{cost, pooled ride-sourcing}}$                                   &            $-0.036^{***}$ &     0.005 &  -6.946 &            $-0.126^{***}$ &     0.018 &  -7.022 \\
\quad $\beta_{\text{time, car}}$                                                    &            $-0.072^{***}$ &     0.005 & -14.724 &            $-0.062^{***}$ &     0.005 & -12.265 \\
\quad $\beta_{\text{time, transit}}$                                                &            $-0.021^{***}$ &     0.003 &  -7.154 &            $-0.024^{***}$ &     0.003 &  -8.084 \\
\quad $\beta_{\text{time, walk}}$                                                   &            $-0.060^{***}$ &     0.002 & -26.591 &            $-0.060^{***}$ &     0.002 & -25.118 \\
\quad $\beta_{\text{time, bike}}$                                                   &            $-0.054^{***}$ &     0.004 & -12.386 &            $-0.056^{***}$ &     0.004 & -12.683 \\
\quad $\beta_{\text{time, taxi, solo ride-sourcing}}$                                &            $-0.043^{***}$ &     0.004 & -12.015 &            $-0.022^{***}$ &     0.005 &  -4.242 \\
\quad $\beta_{\text{time, pooled ride-sourcing}}$                                   &            $-0.040^{***}$ &     0.003 & -13.338 &            $-0.025^{***}$ &     0.004 &  -6.164 \\
\quad $\beta_{\text{park rate at destination, car}}$                                &            $-0.248^{***}$ &     0.006 & -38.267 &            $-0.248^{***}$ &     0.007 & -36.154 \\
\quad $\beta_{\text{no. of transfers, transit}}$                                    &            $-0.148^{***}$ &     0.028 &  -5.192 &            $-0.147^{***}$ &     0.028 &  -5.176 \\
\midrule
Socio-economic attributes & & & & \\
\quad $\beta_{\text{median age at origin, taxi}}$                                   &            $-0.085^{***}$ &     0.016 &  -5.384 &            $-0.084^{***}$ &     0.015 &  -5.474 \\
\quad $\beta_{\text{median age at origin, solo ride-sourcing}}$                     &  $\phantom{-}0.044^{***}$ &     0.014 &   3.195 &  $\phantom{-}0.044^{***}$ &     0.014 &   3.121 \\
\quad $\beta_{\text{median age at origin, pooled ride-sourcing}}$                   &                  $-0.020$ &     0.013 &  -1.494 &                  $-0.010$ &     0.014 &  -0.729 \\
\quad $\beta_{\text{median age squared at origin, taxi}}$                           &  $\phantom{-}0.080^{***}$ &     0.009 &   9.129 &  $\phantom{-}0.082^{***}$ &     0.007 &  11.485 \\
\quad $\beta_{\text{median age squared at origin, solo ride-sourcing}}$             &                  $-0.012$ &     0.009 &  -1.358 &                  $-0.009$ &     0.008 &  -1.102 \\
\quad $\beta_{\text{median age squared at origin, pooled ride-sourcing}}$           &             $-0.021^{**}$ &     0.008 &  -2.534 &            $-0.021^{***}$ &     0.008 &  -2.644 \\
\quad $\beta_{\text{median income at origin, taxi}}$                                &  $\phantom{-}0.576^{***}$ &     0.026 &  22.245 &  $\phantom{-}0.582^{***}$ &     0.027 &  21.918 \\
\quad $\beta_{\text{median income at origin, solo ride-sourcing}}$                  &  $\phantom{-}0.566^{***}$ &     0.022 &  25.581 &  $\phantom{-}0.571^{***}$ &     0.023 &  24.418 \\
\quad $\beta_{\text{median income at origin, pooled ride-sourcing}}$                &  $\phantom{-}0.159^{***}$ &     0.019 &   8.421 &  $\phantom{-}0.151^{***}$ &     0.021 &   7.229 \\
\quad $\beta_{\text{median income squared at origin, taxi}}$                        &            $-0.087^{***}$ &     0.011 &  -8.059 &            $-0.088^{***}$ &     0.010 &  -8.398 \\
\quad $\beta_{\text{median income squared at origin, solo ride-sourcing}}$          &            $-0.101^{***}$ &     0.008 & -12.439 &            $-0.098^{***}$ &     0.010 &  -9.690 \\
\quad $\beta_{\text{median income squared at origin, pooled ride-sourcing}}$        &            $-0.030^{***}$ &     0.008 &  -3.615 &            $-0.026^{***}$ &     0.009 &  -3.027 \\
\quad $\beta_{\text{prop. of households with zero cars at origin, car}}$            &            $-0.436^{***}$ &     0.020 & -22.080 &            $-0.440^{***}$ &     0.018 & -24.733 \\
\midrule
Land use and built environment & & & & \\
\quad $\beta_{\text{residential density at origin, taxi}}$                          &  $\phantom{-}0.413^{***}$ &     0.012 &  33.396 &  $\phantom{-}0.413^{***}$ &     0.011 &  36.021 \\
\quad $\beta_{\text{residential density at origin, solo ride-sourcing}}$            &  $\phantom{-}0.161^{***}$ &     0.010 &  16.260 &  $\phantom{-}0.168^{***}$ &     0.010 &  17.176 \\
\quad $\beta_{\text{employment density at origin, taxi}}$                           &  $\phantom{-}0.068^{***}$ &     0.003 &  20.611 &  $\phantom{-}0.067^{***}$ &     0.003 &  21.385 \\
\quad $\beta_{\text{employment density at origin, solo ride-sourcing}}$             &            $-0.015^{***}$ &     0.003 &  -4.421 &            $-0.013^{***}$ &     0.004 &  -3.654 \\
\quad $\beta_{\text{employment density at origin, pooled ride-sourcing}}$           &            $-0.037^{***}$ &     0.004 & -10.430 &            $-0.038^{***}$ &     0.003 & -11.225 \\
\quad $\beta_{\text{land use diversity at origin, taxi}}$                           &            $-1.378^{***}$ &     0.198 &  -6.950 &            $-1.406^{***}$ &     0.162 &  -8.678 \\
\quad $\beta_{\text{land use diversity at origin, solo ride-sourcing}}$             &            $-0.526^{***}$ &     0.151 &  -3.483 &            $-0.494^{***}$ &     0.136 &  -3.618 \\
\quad $\beta_{\text{land use diversity at origin, pooled ride-sourcing}}$           &            $-0.443^{***}$ &     0.145 &  -3.054 &            $-0.468^{***}$ &     0.123 &  -3.796 \\
\quad $\beta_{\text{pedestrian network density at origin, transit}}$                &            $-0.245^{***}$ &     0.029 &  -8.589 &            $-0.241^{***}$ &     0.031 &  -7.705 \\
\quad $\beta_{\text{pedestrian network density at origin, taxi}}$                   &            $-0.504^{***}$ &     0.026 & -19.694 &            $-0.500^{***}$ &     0.025 & -20.311 \\
\quad $\beta_{\text{pedestrian network density at origin, solo ride-sourcing}}$     &            $-0.240^{***}$ &     0.024 &  -9.809 &            $-0.239^{***}$ &     0.026 &  -9.169 \\
\quad $\beta_{\text{pedestrian network density at origin, pooled ride-sourcing}}$   &            $-0.323^{***}$ &     0.024 & -13.195 &            $-0.333^{***}$ &     0.026 & -12.787 \\
\quad $\beta_{\text{intersection density at origin, transit}}$                      &  $\phantom{-}0.242^{***}$ &     0.021 &  11.456 &  $\phantom{-}0.235^{***}$ &     0.023 &  10.207 \\
\quad $\beta_{\text{intersection density at origin, taxi}}$                         &  $\phantom{-}0.397^{***}$ &     0.021 &  19.069 &  $\phantom{-}0.394^{***}$ &     0.020 &  19.659 \\
\quad $\beta_{\text{intersection density at origin, solo ride-sourcing}}$           &  $\phantom{-}0.145^{***}$ &     0.020 &   7.120 &  $\phantom{-}0.140^{***}$ &     0.022 &   6.433 \\
\quad $\beta_{\text{intersection density at origin, pooled ride-sourcing}}$         &  $\phantom{-}0.192^{***}$ &     0.021 &   9.192 &  $\phantom{-}0.194^{***}$ &     0.020 &   9.587 \\
\midrule
Weather conditions & & & & \\
\quad $\beta_{\text{avg. temperature, taxi}}$                                       &  $\phantom{-}0.162^{***}$ &     0.017 &   9.457 &  $\phantom{-}0.164^{***}$ &     0.020 &   8.363 \\
\quad $\beta_{\text{avg. temperature, solo ride-sourcing}}$                         &  $\phantom{-}0.151^{***}$ &     0.016 &   9.150 &  $\phantom{-}0.150^{***}$ &     0.015 &   9.819 \\
\quad $\beta_{\text{avg. temperature, pooled ride-sourcing}}$                       &  $\phantom{-}0.116^{***}$ &     0.016 &   7.078 &  $\phantom{-}0.107^{***}$ &     0.016 &   6.514 \\
\quad $\beta_{\text{daily precipitation, taxi}}$                                    &  $\phantom{-}0.917^{***}$ &     0.074 &  12.398 &  $\phantom{-}0.917^{***}$ &     0.071 &  12.971 \\
\quad $\beta_{\text{daily precipitation, solo ride-sourcing}}$                      &  $\phantom{-}0.930^{***}$ &     0.073 &  12.750 &  $\phantom{-}0.979^{***}$ &     0.064 &  15.363 \\
\quad $\beta_{\text{daily precipitation, pooled ride-sourcing}}$                    &  $\phantom{-}0.992^{***}$ &     0.067 &  14.888 &  $\phantom{-}1.050^{***}$ &     0.072 &  14.617 \\
\quad $\beta_{\text{avg. temperature x daily precipitation, taxi}}$                 &  $\phantom{-}1.483^{***}$ &     0.133 &  11.182 &  $\phantom{-}1.485^{***}$ &     0.125 &  11.909 \\
\quad $\beta_{\text{avg. temperature x daily precipitation, solo ride-sourcing}}$   &  $\phantom{-}1.568^{***}$ &     0.137 &  11.406 &  $\phantom{-}1.556^{***}$ &     0.124 &  12.533 \\
\quad $\beta_{\text{avg. temperature x daily precipitation, pooled ride-sourcing}}$ &  $\phantom{-}1.691^{***}$ &     0.124 &  13.599 &  $\phantom{-}1.683^{***}$ &     0.117 &  14.437 \\
\midrule
Control function & & & & \\
\quad $\phi_{\text{solo ride-sourcing}}$                                            &                        &           &         &  $\phantom{-}0.060^{***}$ &     0.013 &   4.709 \\
\quad $\phi_{\text{pooled ride-sourcing}}$                                          &                        &           &         &  $\phantom{-}0.094^{***}$ &     0.019 &   4.989 \\
\midrule
\multicolumn{7}{l}{Significance levels: $^{***}$ $p < 0.01$, $^{**}$ $p < 0.05$, $^{*}$ $p < 0.1$} \\
\bottomrule
\end{tabular}

%% file: table_results_first_stage.tex
\begin{tabular}{l|lrr|lrr}
\toprule
{} & \multicolumn{3}{c|}{Price: solo r.-s.} & \multicolumn{3}{c}{Price: pooled r.-s.} \\
{} &  Est. & Std. err. & z-stat. &       Est. & Std. err. & z-stat. \\
\midrule
$\gamma_{\text{driving cost}}$                           &  $\phantom{-}0.206^{***}$ &     0.002 &  82.703 &  $\phantom{-}0.145^{***}$ &     0.002 &  70.331 \\
$\gamma_{\text{transit service frequency}}$              &  $\phantom{-}0.052^{***}$ &     0.008 &   6.465 &            $-0.030^{***}$ &     0.006 &  -4.745 \\
\midrule
\multicolumn{7}{l}{Significance levels: $^{***}$ $p < 0.01$, $^{**}$ $p < 0.05$, $^{*}$ $p < 0.1$} \\
\bottomrule
\end{tabular}

%% file: table_results_first_stage_sum.tex
\begin{tabular}{lrr}
\toprule
{} &        Price: solo r.-s. &  Price: pooled r.-s. \\
\midrule
F-stat.          & 14292.992 &    6429.015 \\
p-val. (F-stat.) &     0.000 &       0.000 \\
R-squared        &     0.812 &       0.661 \\
\bottomrule
\end{tabular}

%% file: table_results_second_stage_sum.tex
\begin{tabular}{lcc}
\toprule
 & MNL uncorrected & MNL with control function \\
\midrule
No. of parameters & 77 & 79 \\
Null log-lik.  & -141259.46 & -141259.46 \\
Log-lik. & -75478.51 & -75405.76 \\
\bottomrule
\end{tabular}

%% file: table_elas_own.tex
\begin{tabular}{lrrrrrrr}
\toprule
{} &    Car &  Transit &   Walk &   Bike &   Taxi &    Solo r.-s. &  Pooled r.-s. \\
\midrule
Uncorrected model \\
\quad Cost &  -0.007 &   -0.050 &    NaN &    NaN & -0.851 & -0.530 &      -0.262 \\
\quad Time & -0.437 &   -0.246 & -1.027 & -1.159 & -0.649 & -0.791 &      -0.951 \\
Corrected model \\
\quad Cost & -0.078 &   -0.067 &    NaN &    NaN & -1.204 & -1.078 &      -0.923 \\
\quad Time & -0.375 &   -0.276 & -1.033 & -1.204 & -0.326 & -0.397 &      -0.600 \\
\bottomrule
\end{tabular}

%% file: table_lit_review.tex
\begin{longtable}{L{3cm} L{2.5cm} L{3cm} L{0.75cm} L{1cm} L{11cm}}
\hline
\textbf{Study} & \textbf{Location} & \textbf{Modelling approach} & \textbf{Data type} & \textbf{Study type} & \textbf{Key findings} \\ \hline
\cite{acheampong2020mobility} & Ghana & Structural equation & 2 & 1, 3 & Perceived benefits, ease of use, perceived safety risks, and car-dependent lifestyles are associated with the adoption and use of ride-hailing services. Ride-hailing is used alone for full door-to-door journeys, instead of complementing other travel modes. \\ \hline
\cite{alonso2020value} & Netherlands & Mixed and latent class logit & 3 & 4 & Value of in-vehicle travel time for pooled on-demand services is 7.88–10.80 euro per hour. Value of reliability is around half of the value of wait time and in-vehicle travel time. \\ \hline
\cite{asgari2020incorporating} & USA & Error component logit & 3 & 3, 4, 5 & This study creates a habit index based on past usage frequency of a mode. They find that habits and private vehicle expenses (e.g., parking cost and time spent in finding it) are highly associated with the preferences towards rider-sourcing services. \\ \hline
\cite{azimi2021exploring} & USA & Error component nested logit & 3 & 3, 4, 5 & Generation Xers and Millennials have distinct preferences for on-demand shared mobility.  Whereas the perceived time and cost benefits of shared mobility affect Generation Xers’ preferences for shared mobility, Millennials’ choices are more likely to be influenced by their attitudes towards on-demand services. \\ \hline
\cite{baker2020transportation} & San Francisco, USA & Geographically weighted regression & 1 & 1, 2 & This study finds positive relationships between ride-sourcing use and public transport ridership, as well as between ride-sourcing demand and choice rider (i.e., not transit dependent) neighbourhoods. \\ \hline
\cite{bansal2020eliciting} & USA & Multinomial logit & 2 & 3, 4 & 10\% of ride-sourcing users in the USA postponed the purchase of a new car due to availability of ride-sourcing services. Older ride-sourcing users with higher vehicle ownership are less likely to pool rides. Ride-sourcing drivers with a postgraduate degree who drive daily and live in metropolitan regions are more likely to switch to fuel-efficient vehicles. \\ \hline
\cite{bi2021exploring} & Chengdu, China & Latent dirichlet allocation & 1 & 1, 2 & Ride-sourcing in Chengdu is mostly used for non-work trips. Ride-sourcing is more extensively used in areas with lack of public transit access. \\ \hline
\cite{dean2021spatial} & Chicago, USA & Poisson-Gamma negative binomial and linear regression & 1 & 2, 4 & Longer trips in Chicago are likely to be requested as shared rides. Census tracts with higher shares of young, unemployed, and non-White persons, and vehicle-free households have higher proportions of shared rides. Shared ride-hail demand decreases with the increase in density of pedestrian infrastructure. \\ \hline
\cite{dey2021transformation} & New York City, USA & Negative Binomial and fractional split & 1 & 1, 2 & Job density, employment density, bike infrastructure and transit service significantly affect ride-hailing demand. This study also provides a prediction framework for predicting future ride-hailing trends. \\ \hline
\cite{dong2021impact} & Boston and Philadelphia, USA & Mixed logit & 3 & 1, 3, 4, 5 & Carless households are likely to delay or forgo car purchasing decisions due to ride-sourcing services. TNCs are likely to substitute transit more than complementing it. Willingness to pay to save 10 minutes of wait time for transit is US\$5 in Boston and US\$8 in Philadelphia, compared to below US\$3 and US\$2 for TNCs. \\ \hline
\cite{dong2020trade} & Philadelphia, USA & Mixed logit & 3 & 1, 3, 5 & Higher-income females with age over 30 years, who are less frequent transit users, are increasingly willing to choose ride-hailing over transit. Time spent on walking to transit stop is found to be more burdensome than in-vehicle travel time and wait time. \\ \hline
\cite{edwards2020nonstandard} & Austin, USA & Linear regression & 1 & 2 & Nonstandard (expansive variants) ride-sourcing services are preferable by airport travellers and those living in carless and low-density neighbourhoods, but are less preferred in low-income neighbourhoods. \\ \hline
\cite{ghaffar2020modeling} & Chicago, USA & Random-effects negative binomial regression & 1 & 1, 2 & Higher ride-sourcing demand is experienced in Chicago on days with lower temperature. Census tracts with higher household income, high proportion of carless households, higher employment and population density, and fewer parking spots have higher demand for ride-sourcing services. \\ \hline
\cite{gomez2021adoption} & Madrid, Spain & Generalized heterogeneous data & 2 & 1, 3 & Young, well-educated, wealthy individuals, who are familiar with new technologies, have more inclination towards ride-hailing services. Environment-conscious respondents are less inclined towards ride-hailing. Ride-hailing has substituted transit for leisure and errand trips. \\ \hline
\cite{hasnine2021effects} & Toronto, Canada & Autoregressive moving average & 1 & 1, 2 & Lagged demand is a good predictor of the future demand for ride-sourcing services. Households that rent their dwelling have positive association with ride-sourcing demand. Bike-sharing and transit trip counts are negatively correlated with the ride-sourcing demand, but precipitation has a positive relationship. \\ \hline
\cite{hou2020factors} & Chicago, USA & Linear regression and machine-learning & 1 & 2,4 & Ride-sourcing trips starting or ending at Chicago-area airports have a smaller ratio for shared trips, perhaps due to time, luggage, and other constraints. \\ \hline
\cite{kang2021pooled} & Austin, USA & Generalized heterogeneous data & 3 & 3, 4, 5 & Women, older adults, and non-Hispanic/non-Latino Whites have a low propensity to pool rides in Austin. Austin residents are willingness to pay to not pool a ride is on averages about 62 cents for commute, US\$1.70 for shopping and US\$1.32 for leisure travel. \\ \hline
\cite{lazarus2021pool} & California, USA & Multinomial logit & 3 & 3, 4, 5 & High-frequency ride-sourcing users are more likely to share rides. Average values of in-vehicle travel time for Los Angeles, Sacramento, San Diego, and San Francisco are US\$29.2, US\$27.3, US\$25.9, and US\$34.5. \\ \hline
\cite{li2021exploring} & Toronto, Canada & Random-effects panel data and log–log regression & 1 & 2 & Ride-sourcing and public transit demand varies by transit mode, time of day and transit level-of-service. The demand for ride-sourcing services is positively associated with subway station ridership during the mid-day and early evening, while negatively correlating with surface transit demand during peak commuting hours. \\ \hline
\cite{loa2021influences} & Toronto, Canada & Structural equation & 2 & 1, 3 & Students, persons from lower-income households, and transit pass owners are more likely to substitute ride-sourcing for public transit. \\ \hline
\cite{loa2021examining} & Toronto, Canada & Binary logistic and zero-inflated ordered probit & 2 & 3, 4 & Transit pass ownership positively influences the frequency of ride-hailing usage (i.e., complementary relationship). The factors affecting preference for shared and single-occupancy ride-hailing differ, and therefore, both modes need to be studied separately. \\ \hline
\cite{malik2021deeper} & California, USA & Integrated choice and latent variable & 2 & 1, 3 & Omission of variables related to residential location and vehicle ownership could bias the results on the linkage between transit ridership and ride-hailing. To discourage the replacement of active mode through ride-hailing services, pricing strategies should be employed to reduce the use of ride-hailing for short trips. \\ \hline
\cite{marquet2020spatial} & Chicago, USA & Truncated Poisson & 1 & 1, 2 & Walkable and diverse neighbourhoods attract and generate more ride-sourcing trips. Areas with lower car ownership generate fewer ride-sourcing trips but attract more trips. \\ \hline
\cite{nugroho2020explaining} & Indonesia & Ordered logit & 2 & 1, 3 & Lack of car access encourages the adoption of ride-sourcing services. Ride-sourcing services complement the local public transport system. \\ \hline
\cite{sabogal2021not} & Mexico city & Multinomial and ordered logit & 2 & 3 & Ride-hailing is mainly used for leisure and health trips. Young travellers with higher education and higher income are more inclined to adopt ride-hailing. Due to perception of crime and sexual harassment in Mexico city, women depend more on ride-sourcing than men. \\ \hline
\cite{sabouri2020exploring} & USA & Multi-level linear & 1 & 1, 2 & Census block level population, employment, activity, and transit density are positively associated with the Uber demand, but intersection density and destination accessibility by auto and transit has negative effect. \\ \hline
\cite{shen2020modeling} & Nanjing, China & Nested logit & 3 & 3, 5 & Young travellers are naturally inclined to use ride-sourcing, but age does not make difference in preference of the service type (premier vs. regular). The demand for ride-sourcing services is elastic to in-vehicle travel time and waiting time, but much highly elastic to travel cost. \\ \hline
\cite{soltani2021ridesharing} & Adelaide, Australia & Multinomial logit & 2 & 2, 3 & Younger travellers with higher levels of education and income, who live in dense areas with higher property prices, are more inclined towards ride-hailing services. Car ownership, ethnic background, gender, and household size have no association with the propensity to use ride-hailing services. \\ \hline
\cite{sweet2021user} & Ontario, Canada & Mixed logit & 3 & 3, 4, 5 & An individual is willing to pay US\$1 to US\$4 for not sharing trip with other passenger. The analysis provides mixed evidence about the potential of integrating transit with ride-sourcing to solve the last mile problem. \\ \hline
\cite{toman2020dynamic} & New York City, USA & Multivariate time series & 1 & 1 & This study explored substitution effects between ride-sourcing, shared bike, taxi, and subway. The relationship varies across weekdays and weekends, and also during holidays. \\ \hline
\cite{tu2021exploring} & Chengdu, China & Gradient boosting decision trees & 1 & 2,  4 & Distance to city centre, land use diversity and road density are the main determinants of the proportion of shared ride-sourcing trips. Whereas the first two factors positively affect the proportion of shared trips, the last one and public transport density has negative affect. \\ \hline
\cite{von2021exploring} & China & Integrated choice and latent variable & 2 & 1, 3 & Ride-hailing services are more attractive in tier-2 cities due to poor public transport infrastructure. Ride-hailing substitutes other modes for medium-distance trips (5-10 km). Women with higher income are more likely to use ride-sourcing services than men. \\ \hline
\cite{ward2021impact} & USA & Difference-in-difference & 1 & 1 & Entry of ride-sourcing services led to an increase of 0.7\% in vehicle registrations. The effect varies across urban areas. There is no significant effect of the entry of ride-souring services on the transit use. \\ \hline
\cite{wang2022identifying} & Michigan, USA & Latent class cluster analysis & 2 & 3, 4 & Males, college graduates, and car owners have a higher inclination toward ride-sourcing services. Vehicle owners have lesser interest in sharing ride-sourcing trips than their demographic counterparts. \\ \hline
\cite{yan2020using} & Chicago, USA & Random forest & 1 & 2 & Among the built-environment variables, employment density and walkability at trip origin have high correlation with ride-sourcing demand. Among the transit-supply factors, frequencies of bus and rail services have the strongest correlation with ride-sourcing demand. \\ \hline
\cite{yan2021mobility} & Michigan, USA & Ordered logit & 2 & 1, 3 & Male college graduates with poor transit access are more inclined to use mobility-on-demand transit services, but people with lack of access to mobile data have significantly lower preference for such services. Disability does not seem to be associated with the preference for mobility-on-demand services. \\ \hline
\cite{yu2020impacts} & Austin, Texas & Structural equation & 1 & 2 & Population/employment/road density, and transit accessibility have positive association with ride-sourcing demand, but walk accessibility have a negative effect. These effects vary across the time-of-day. \\ \hline
\multicolumn{6}{l}{\textbf{Note 1:} In column ``data type", 1: spatial trip level, census, weather, land use data, 2: household travel survey, 3: discrete choice experiment.} \\
\multicolumn{6}{l}{\begin{tabular}[c]{@{}l@{}}\textbf{Note 2:} In column ``study type", 1: substitution/complementary effect of ride-sourcing on other travel modes and vehicle ownership, 2: correlation between a spatial unit's characteristics \\ (e.g., population density) and ride-sourcing demand, 3: relation of individual-level attitudes and socio-economic attributes with preference for ride-sourcing services,4: factors affecting \\ preference   for shared ride-sourcing services  5: effect of mode-specific attributes on the demand for ride-sourcing services.\end{tabular}} \\
\caption{Overview of recently published ride-sourcing demand analysis studies}
\label{tab:litreview}
\end{longtable}